\documentclass[traditabstract]{aa}
\usepackage{graphicx}
\usepackage{times,epsfig,amssymb}
\usepackage{txfonts}
\usepackage{natbib}
\usepackage{color}
\begin{document}

\authorrunning{S. De Grandi et al.}
\titlerunning{On the Fe abundance  peak formation in cool-core clusters of galaxies}

  \title{On the Fe abundance  peak formation in cool-core clusters of galaxies:
   hints from cluster WARPJ1415.1+3612 at z=1.03}

 %  \subtitle{sottotitolo}

   \author{Sabrina De Grandi \inst{1}
          \and
          Joana S. Santos\inst{2}
          \and
          Mario Nonino\inst{3}
           \and
          Silvano Molendi\inst{4}
          \and
          Paolo Tozzi\inst{3}
         \and
          Mariachiara Rossetti\inst{5}
          \and %\\
          Alexander Fritz\inst{4}
          \and
          Piero Rosati\inst{6}
}

%    \institute{
   \institute{INAF, Osservatorio Astronomico di Brera,
              via E. Bianchi 46, I-23807 Merate, Italy\\
             \email{sabrina.degrandi@brera.inaf.it}
         \and
             INAF, Osservatorio Astrofisico di Arcetri,
             Largo Enrico Fermi 5, I-50125 Firenze, Italy\\
             \email{jsantos@arcetri.astro.it,}
             \email{ptozzi@arcetri.astro.it}
         \and
             INAF, Osservatorio Astronomico di Trieste,
             via G.B. Tiepolo 11, I-34143 Trieste, Italy\\
             \email{nonino@oats.inaf.it}
         \and
             INAF - IASF, via Bassini 15, I-20100 Milano, Italy\\
             \email{silvano@iasf-milano.inaf.it}
         \and
             Universit\`a di Milano, I-20100 Milano, Italy\\
             \email{mariachiara.rossetti@unimi.it}
         \and
             Universit\`a di Ferrara, Via Saragat 1, I-44122 Ferrara, Italy\\
             \email{rosati@fe.infn.it}
}

   \date{Received April 2, 2014; accepted May 30, 2014}
%   \date{}

% \abstract{}{}{}{}{}
% 5 {} token are mandatory

 \abstract
{We present a detailed study of the iron content of the core of the high-redshift cluster WARPJ1415.1+3612 ($z=1.03$).
By comparing the central Fe mass excess observed in this system,
$M_{\rm Fe}^{\rm exc} = (1.67\pm0.40)\times 10^9$ M$_\odot$,
with those measured in local cool-core systems, we infer that the bulk of
the mass excess was already in place at $z=1$,
when the age of the Universe was about half of what it is today.
Our measures point to an early and intense period of star formation
most likely associated with the formation of the BCG.
Indeed, in the case of the power-law delay time distribution with slope $-1$, 
which reproduces the data of WARPJ1415.1+3612  best, half of the 
supernovae explode within 0.4 Gyr of the formation of the BCG.
Finally, while for local cool-core clusters the Fe distribution is
broader than the near infrared light distribution of the BCG, in
WARPJ1415.1+3612 the two distributions are consistent, indicating
that the process responsible for broadening the Fe distribution in
local systems has not yet started in this distant cluster.
}

% context heading (optional) leave it empty if necessary


   \maketitle
%
%________________________________________________________________

\section{Introduction}

The deep potential wells of clusters of galaxies are permeated by a
hot (kT $\sim 1-10$ keV), optically thin and spatially diffused plasma,
the intra-cluster medium (ICM), which is also enriched by heavy elements.
It follows that observations in the X-ray
band provide a powerful method for studying and classifying these systems.
Galaxy clusters can be divided into two main classes, cool-core and non
cool-core systems, on the basis of the thermodynamical and chemical
properties of their central regions (i.e. within radii $\simeq$ 0.15
r$_{500}$).
Cool-core clusters show an almost regular, elliptical morphology on
large scales \citep[e.g.][]{nurgaliev13}, and their innermost region displays
several
characteristic properties: the surface brightness and the iron (Fe)
abundance are both strongly peaked at the centre
(where central cooling times are significantly shorter than the Hubble time)
\citep[e.g.][]{santos08,sanderson10},
the temperature decreases to about one third of the external value
\citep[e.g.][]{hudson10,panagoulia14},
and the central entropy is $\lesssim 30$ keV cm$^2$
\citep[e.g.][]{cavagnolo09,pratt10,mcdonald13}.
Conversely, non-cool-core clusters do not show prominent
surface-brightness peaks and temperature drops in the cores, and very
often, they feature the irregular X-ray morphologies and galaxy distributions
indicative of recent merging events.

The study of the spatial distribution of iron in the ICM
provides important information that can be used to deepen our
understanding of the chemical history and evolution of galaxy clusters
in general and of cool-core clusters in particular.
Outside the core, the Fe abundance of the ICM is on average one-fourth
solar for both cool-core and non-cool-core clusters
(\citealt[][hereafter DM01;]{degrandi01_ab} \citealt{leccardi10,matsushita11}),
and it shows only a moderate
evolution up to a redshift of $\sim1.3$
\citep{balestra07,maughan08,baldi12}.

The abundance excess found at the centre of cool core clusters is always
associated with a giant, early-type galaxy, the brightest cluster galaxy (BCG),
often showing a very extended diffuse stellar envelope.
The favoured scenario for the formation of the BCGs is through mergers
of galaxies falling in along primordial filaments early in the history
of the Universe
\citep[e.g.][]{dubinsky98,delucia07,whiley08,collins09,hung12,
ruszkowski09,lin13,lidman13}.
The presence of a BCG makes this galaxy the prime candidate for
producing the peaked abundance profiles
\citep[e.g.][]{fukazawa00,boehringer04b,degrandi04,leccardi08_ab}.
\citet{degrandi04} estimated that BCGs in low-redshift clusters (i.e. 
$z<0.1$) are able to produce the Fe mass observed in the cluster core
during their star formation histories.  Furthermore, the observations
show that the stellar light profiles of the BCGs are
less extended than the Fe abundance profiles, suggesting that there
must be a mixing mechanism that moved the metals
from their production sites
\citep[DM01;][]{churazov03,rebusco05,graham06}.

Very little is known about the formation time and subsequent evolution
of these peaked Fe abundance profiles in cool-core clusters.
\citet{baldi12} present the first attempt to determine, with
XMM-Newton and spatially resolved data, the evolution of the abundance at
different positions in the clusters and with redshift (up to $z\sim1.4$) and
find no evidence of any evolution in the Fe content in the cores,
although with low statistical significance and without distinguishing
cool cores from non-cool cores.
The major sources of Fe in the Universe are type-Ia supernovae (SNIa),
whereas other elements (alpha elements) are produced by core-collapsed
SN (SNcc) or by a mix of the two SN types. A measure of the
alpha/Fe ratios in the ICM gives a 
first-order approximation of the relative importance of the SN types to
the enrichment \citep[for a review see][]{werner08_rev,borgani08_reva}.
Early claims of an increasing type-Ia supernovae (SNIa)
contribution, with respect to the SN core-collapsed (SNcc)
to the ICM enrichment towards cluster centres \citep[][]
{finoguenov00,rasmussen07}, contributed to establishing a
scenario where the Fe abundance in cores increased steadily from
the time of formation to $z=0$ \citep[e.g.][]{fukazawa00,boehringer04b}.
More recent observations from XMM-Newton, Chandra, and SUZAKU
\citep[e.g.][]{tamura04,million11,sato07_sn,matsushita13},
which found a constant ratio between
SN types at any radius in cool-core clusters contradicting the
early results, requires to reconsider this picture.

In this paper we provide a detailed investigation of  iron
content and spatial distribution in the X-ray-selected cluster
WARPJ1415.1+3612 (hereafter WARPJ1415) at redshift $z = 1.03$
\citep{perlman02,fritz09}, using a deep Chandra observation recently
explored by \citet[hereafter S12]{santos12}.
This cluster is a high-redshift, strong cool-core cluster \citep[S12,][]{santos10}, which
is strikingly similar in its thermodynamic characteristics to nearby
cool-core clusters.  In particular we study the connection between its
Fe content and its central galaxy,
which was observed with the Subaru and Spitzer telescopes.
Our aim is to compare the properties of this distant cluster with
those of local (i.e. z $<0.1$) cool-core systems
to try to constrain the formation history of their iron peaks \citep[DM01;][]{degrandi04}.
We extended the analysis in DM01 with XMM-Newton data, selecting from
the XMM-Newton public archive a small sample of nearby clusters
for which we also have near infrared (NIR) observations of the central
BCGs, in order to compare their chemical properties with those of the
distant cluster WARPJ1415.

The structure of this paper is as follows.
In Section 2, we present the X-ray and near-infrared (NIR)
datasets and the data analysis procedures
used to  study  WARPJ1415 and the nearby
cool-core clusters.
In Section 3, we measure the Fe content in WARPJ1415
and in the local clusters and compare them. We estimate the enrichment
times and the supernovae rates needed to produce the observed Fe
mass in WARPJ1415.
Section 4 is devoted to comparing the BCG's star-light
profiles and the Fe abundance distributions in the ICM in our distant
and local clusters and to discussing the ensuing implications for the
enrichment histories in cool-core cluster cores.
Finally, we summarize our results in Section 5.

Throughout this paper we assume the concordance  flat $\Lambda$CDM
cosmology with $H_0 = 71$ km s$^{-1}$
Mpc$^{-1}$, $\Omega_{\rm m} = 0.27,$ and $\Omega_\Lambda = 0.73$,
that for WARPJ1415 at $z=1.03$ gives a scale of 8.099 kpc/$^{\prime\prime}$.
Iron abundances are measured relative to the solar photospheric
values of \citet{asplund09}, where the solar Fe abundance is
$Z_{\rm Fe,\odot} = 3.16\times 10^{-5}$.

Quoted errors are at the $1\sigma$ level, unless otherwise stated.

%__________________________________________________________________

\section{Data and analysis}

\subsection{WARPJ1415}

\subsubsection{X-ray data}

The galaxy cluster WARPJ1415 was detected in the Wide Angle ROSAT
Pointed Survey \citep{jones98,perlman02} and then
observed by Chandra ACIS-S (280 ks) and ACIS-I (90 ks).
S12 performed a detailed spatially resolved spectral analysis using
both Chandra datasets for a total of 370 ks exposure time.
Their analysis so far represents the most detailed X-ray analysis of any
cool-core cluster at such high redshift, $z\sim 1.$.
Details on the data reduction with the CIAO 4.3 software package with
CALDB 4.4.5 \citep{fruscione06} are given in Sect. 2 of S12.
%
%----------------------------------------------------------------------
\begin{table}
\begin{center}
\caption{ Best-fitting values for a 2-temperature {\it apec} model for
  the ACIS (S and I) spectrum of the innermost ring.  The columns are:
  (1) first component temperature that is fixed, (2) best-fit of the
  second component temperature, (3) ratio between the normalizations
  of the first and second components, (4) best fit of the Fe abundance
  in solar units, (5) Cash statistics of the best fit (always for 251
  PHA bins).  The last row shows the result for the single temperature
  fit of the spectrum.
}
\label{tab_febias}
\begin{tabular}{c c c c c}                                                % centered columns (4 columns)
& & & & \\
\hline\hline                                                                  % inserts double horizontal lines
T1 fixed & T2 free                           & N1/N2 & $Z_{\rm Fe}$ & C-stat \\       % table heading
\hline
% apec and ASPL09 S+I ring0 251 pha, files xcm: 
 0.3        &  4.94$^{+0.67}_{-0.56}$ &  $0.261$  &  3.61$^{+1.59}_{-1.06}$  &  199.82 \\
 0.5        &  4.98$^{+0.73}_{-0.58}$ &  $0.094$  &  3.69$^{+1.63}_{-1.09}$  &  200.01 \\
 0.8        &  4.95$^{+0.78}_{-0.60}$ &  $0.052$  &  3.78$^{+1.67}_{-1.12}$  &  200.56 \\
 1.0        &  4.97$^{+0.81}_{-0.62}$ &  $0.050$  &  3.78$^{+1.66}_{-1.12}$  &  200.57 \\
 1.5        &  5.11$^{+0.95}_{-0.74}$ &  $0.085$  &  3.87$^{+1.70}_{-1.13}$  &  200.72 \\
 2.0        &  5.17$^{+2.18}_{-1.20}$ &  $0.147$  &  4.05$^{+3.12}_{-1.65}$  &  201.20 \\
 2.5        &  4.91$^{+3.54}_{-0.97}$ &  $0.134$  &  4.21$^{+3.96}_{-1.78}$  &  201.65 \\ 
 ---         &  4.44$^{+0.46}_{-0.36}$ &  ---  &  4.01$^{+0.99}_{-1.19}$  &  202.00 \\ 
 % apec and ASPL09 only S 176 pha
% 0.1        &  4.32$^{+0.53}_{-0.39}$ &  43.86  &  4.40$^{+2.62}_{-1.41}$  &  135.7973 \\
% 0.3        &  4.37$^{+0.61}_{-0.41}$ &  0.167  &  4.53$^{+2.58}_{-1.46}$  &  135.71 \\
% 0.5        &  4.38$^{+0.66}_{-0.43}$ &  0.059  &  4.63$^{+2.61}_{-1.51}$  &  135.81 \\
% 0.8        &  4.36$^{+0.68}_{-0.48}$ &  0.032  &  4.70$^{+2.76}_{-1.52}$  &  136.12 \\
% 1.0        &  4.41$^{+0.70}_{-0.49}$ &  0.030  &  4.68$^{+2.75}_{-1.52}$  &  136.15 \\
% 1.5        &  4.36$^{+0.82}_{-0.60}$ &  0.052  &  4.70$^{+2.62}_{-1.52}$  &  136.25 \\
% 2.0        &  4.48$^{+1.10}_{-0.65}$ &  0.089  &  4.92$^{+2.74}_{-1.59}$  &  136.44 \\
% 2.5        &  4.53$^{+1.55}_{-0.71}$ &  0.161  &  5.03$^{+2.88}_{-1.65}$  &  136.51 \\
% ---         &  4.10$^{+0.37}_{-0.30}$ &  0.000  &  5.07$^{+3.03}_{-1.69}$  &  136.58 \\
% mekal and AG89
% 0.1        &  4.51$^{+0.53}_{-0.44}$ &  $>$1e-07  &  2.49$^{+0.29}_{-0.37}$  &  124.9376 \\
% 0.3        &  4.68$^{+0.61}_{-0.72}$ &  0.3513      &  2.47$^{+0.38}_{-0.36}$  &  124.7432 \\
% 0.5        &  4.65$^{+0.59}_{-0.41}$ &  0.0932      &  2.51$^{+0.29}_{-0.27}$  &  124.7684 \\
% 0.8        &  4.53$^{+0.74}_{-0.47}$ &  0.0063      &  2.56$^{+0.87}_{-0.62}$  &  124.9335 \\
% 1.0        &  4.53$^{+0.67}_{-0.47}$ &  0.0047      &  2.56$^{+0.73}_{-0.61}$  &  124.9494 \\
% 1.5        &  4.52$^{+0.62}_{-0.46}$ &  4.e-07       &  2.54$^{+0.38}_{-0.59}$  &  124.9442 \\
% 2.0        &  4.51$^{+0.50}_{-0.42}$ &  $>$1e-07  &  2.52$^{+0.27}_{-0.28}$  &  124.9375 \\
% 2.5        &  4.51$^{+0.73}_{-0.47}$ &  $>$1e-07  &  2.52$^{+0.62}_{-0.58}$  &  124.9382 \\
\hline
\end{tabular}

\end{center}
\end{table}
%----------------------------------------------------------------------
%
In S12 the authors extracted ACIS-I and ACIS-S spectra from nine rings
at different radii up to $\sim 400$ kpc and performed the spectral
analysis on the combined spectra (apart for the last ring, 300--400
kpc, where there was no useful signal in the ACIS-I data) by assuming a
single-temperature model.
The resulting projected and de-projected temperature and Fe abundance
profiles are shown in their Fig. 2. (Details on spectral analysis and
de-projection technique are in S12.)

The high X-ray count statistic of WARPJ1415 allows the most detailed
analysis of the Fe abundance distribution in the core of a cluster at
$z\sim 1$ to date.
We performed a new spectral analysis with the {\it XSPEC} package
\citep[v12.6,][]{arnaud96} starting from the same spectra, response
matrices, and the ancillary response matrices computed by S12, but
using the {\it apec} spectral model instead of the {\it mekal} one.
We decided to use {\it apec} because this model is continuously updated
within {\it XSPEC} and we want to be consistent with the analysis of
the XMM-Newton data of the local clusters (see Sect. 2.2.1).  We first
checked that our re-analysis agrees with those published in S12.  To this
aim we combined the ACIS-S and ACIS-I spectra and assumed a single-temperature
 {\it apec} model over the energy range 0.5--7.0 keV and a
fixed local absorption equal to the Galactic neutral hydrogen column
density ($N_H$) measured at the cluster position \citep{kaberla05} of
$N_H = 1.05\times 10^{20}$ cm$^{-2}$.  The gas temperature, metal
abundance, and normalization were left unconstrained in the fitting
procedure.  The resulting projected and de-projected temperature and
Fe abundance profiles are consistent within the uncertainties with
those published in S12.

We tested whether the Fe abundance might be affected by biases induced
by the presence of multi-phase gas in the cluster core.  We do not
expect  the abundance measurement to be strongly affected by
the Fe-bias \citep{buote00b,buote00a} or by the inverse Fe-bias
\citep{rasia08,simionescu09,gasta10} since the
measured emission-weighted temperature in the innermost bin
is $> 4$ keV, and the Fe L-shell complex (energy at rest $\sim$
0.7--1.5 keV) for this cluster at $z=1$ only marginally overlaps the energy range considered for the spectral analysis
($\sim 0.35-0.7$ keV observer frame for a $z=1$ cluster).
We checked the stability of the Fe abundance measurement by fitting
the combined ACIS-S (495 cts) and ACIS-I (94 cts) spectrum of the
innermost bin, a circle with a 20 kpc radius, with a two-temperature
{\it apec}  model, where the temperature of the first
component is fixed and the temperature of the second component is left free
to vary.  The values assigned to the first temperature are reported in
Table \ref{tab_febias}.  As in S12 we use the Cash statistics, which
is preferable for low S/N spectra \citep*{nousek89}.  From the results
of the best-fit shown in Table \ref{tab_febias} it is clear that, if
present, a second component provides a modest contribution to the
total emission and that the value of the measured iron abundance is
essentially unaffected by the multi-temperature structure.

%-----------------------------------------------------------
   \begin{figure}\centering
   \includegraphics[width=6.0cm,angle=-90]{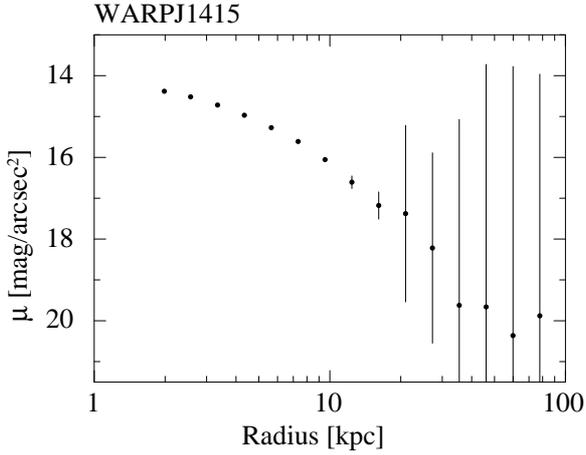}
      \caption{Subaru/MOIRCS Ks surface brightness profile of the BCG
        in WARPJ1415.  }
         \label{fig_optprof_bcg}
   \end{figure}
%-----------------------------------------------------------

\subsubsection{Near-IR data}

The brightest cluster galaxy (BCG) of WARPJ1415 is a large and
massive elliptical galaxy with luminosity in the B band of
$L_B=3\times10^{11} L_{B,\odot}$.
We retrieved the raw broadband images taken with Subaru/Suprime (BVRi'z' bands)
\citep{miyazaki02} and Subaru/MOIRCS (J,Ks bands) \citep[][]{Suzuki08},
from the Subaru archive, SMOKA
\footnote{Based in part on data collected at Subaru Telescope and
obtained from the SMOKA, Subaru-Mitaka-Okayama-Kiso Archive,
which is operated by the Astronomy Data Centre, National Astronomical
Observatory of Japan \citep{baba02}}.
MOIRCS provides wide-field imaging in a $4^\prime\times7^\prime$ field of view
 covered by two $2048\times2048$ arrays with the spatial resolution of
 0.117 arcsec/pixel.
The radial stellar profile in the NIR Ks band is extracted from
MOIRCS data using the IRAF ELLIPSE routine, a procedure that
measures best-fit isophotes from the data.
The sky-background-subtracted profile shown in
Fig. \ref{fig_optprof_bcg} is traced out to a radius of $\sim 80$
kpc; however, beyond a radius of $\sim 30$ kpc, the profile rapidly
dilutes into the background.

Using our own multi-colour catalogue from  broadband optical, infrared,
and IRAC \footnote{This work is based in part on observations made with the
Spitzer Space Telescope obtained from the NASA/IPAC  Infrared Science Archive,
both of which are operated by the Jet Propulsion Laboratory, California
Institute of Technology, under a contract with the National Aeronautics
and Space Administration}
photometry, we estimated the stellar mass
of the BCG by fitting the galaxy spectral energy distribution (SED) with
LePhare \citep{arnouts11} obtaining
$M_{\star} = 9.3\times 10^{11} M_{\sun}$ for a Chabrier IMF
\citep{chabrier03}.%, which is consistent with the dynamical mass by \citet{fritz09}.
With the SED fitting we measured an apparent, rest frame Ks magnitude
of 18.69 AB.
To estimate the Ks rest frame magnitude, we used all photometric
points computed within $1.5^{\prime\prime}$ radius aperture, then corrected
to $2.3^{\prime\prime}$ (18 kpc rest frame) aperture assuming no
colour dependence, with respect to the z band.
This step was adopted to avoid BCG flux contamination in IRAC
bands from nearby sources.
The estimated absolute Ks rest frame magnitude is $-24.73\pm 0.20$ AB.

%----------------------------------------------------------------------
 \begin{table}
 \begin{center}
 \caption{ The local sample of galaxy clusters collected as a
   reference sample in this work.  The columns read as follows: (1)
   name, (2) redshift, (3) XMM-Newton archival observation number and (4)
   effective exposure time for each EPIC detector.  Multiple
   observations of the same cluster were summed up for the spectral
   analysis.}
 \label{tab_loc}
 \begin{tabular}{l c c l c c}
& & & & \\
\hline\hline              
%Name            &    RA  & DEC  & z & XMM ObsId & $t_{exp}$  \\
%                     &    deg & deg  &    &                     &  ks  \\
Name                &  z & Obs. Id & $t_{\rm M1}$ & $t_{\rm M2}$ & $t_{\rm pn}$  \\
                         &     &                     &  ks &  ks &  ks             \\
\hline
A478             &    0.0329    & 0109880101   &    ~39.7 & 92.0 & 61.8   \\
A496             &    0.0329    & 0506260301   &    ~57.1 & 58.7 & 28.3   \\
                     &                   & 0506260401   &    ~53.0 & 47.2 & 30.8   \\
%A2029        &    0.0773    & 0111270101   &       \\
A2597           &    0.0852    & 0147330101   &    ~48.6 & 48.4 & 49.4   \\
A3526           &    0.0114    & 0406200101   &    104.8 & 98.5 & 91.9  \\
%A4038        &    0.0300    & 0204460101   &        \\
A4059           &    0.0475    & 0109950101   &    ~16.3 & 16.1 & 16.1   \\
                     &                   & 0109950201   &    ~22.8 &  22.9 & 19.3  \\
Hydra-A        &    0.0539    & 0504260101   &    ~69.2 & 71.5 & 49.4  \\
%2A0335+096 &    0.0347    & &       &     &     \\
\hline
\end{tabular}

 \end{center}
 \end{table}
%----------------------------------------------------------------------

\subsection{Local clusters of galaxies}

From its thermodynamical properties, the distant cluster WARPJ1415
is very similar to cool-core clusters in the nearby Universe.
In this work, our aim is to study the chemical properties of WARPJ1415
by comparing it with those of local systems. To this purpose we selected,
from the XMM-Newton public archive,
six strong cool-core clusters at redshifts lower than 0.09 that have
medium--high total masses
\citep[i.e. M$_{500}\gtrsim 3.5\times 10^{14} M_\odot$,][]{chen07},
similar to the total mass of WARPJ1415
($M_{500}=2.4\times 10^{14} M_{\odot}$, S12).
In particular, we included in our  sample A3526 (the Centaurus
cluster) since its iron abundance profile is very peaked and
reaches over-solar Fe abundance in the center, as in the case of
WARPJ1415.
We point out that the spatial resolution
for these low-redshift clusters sample observed with XMM-Newton is
very similar to the spatial resolution for a cluster observed
with Chandra at $z\sim1.$ This follows from the fact that the ratio
of the angular distances computed at $z=0.05$ and $z=1$
is similar to the ratio between the point spread functions of the
XMM-Newton/EPIC and Chandra/ACIS instruments.

The local clusters are listed in Table \ref{tab_loc}.

%-----------------------------------------------------------
  \begin{figure*}
   \centering
   \includegraphics[width=6cm,angle=-90]{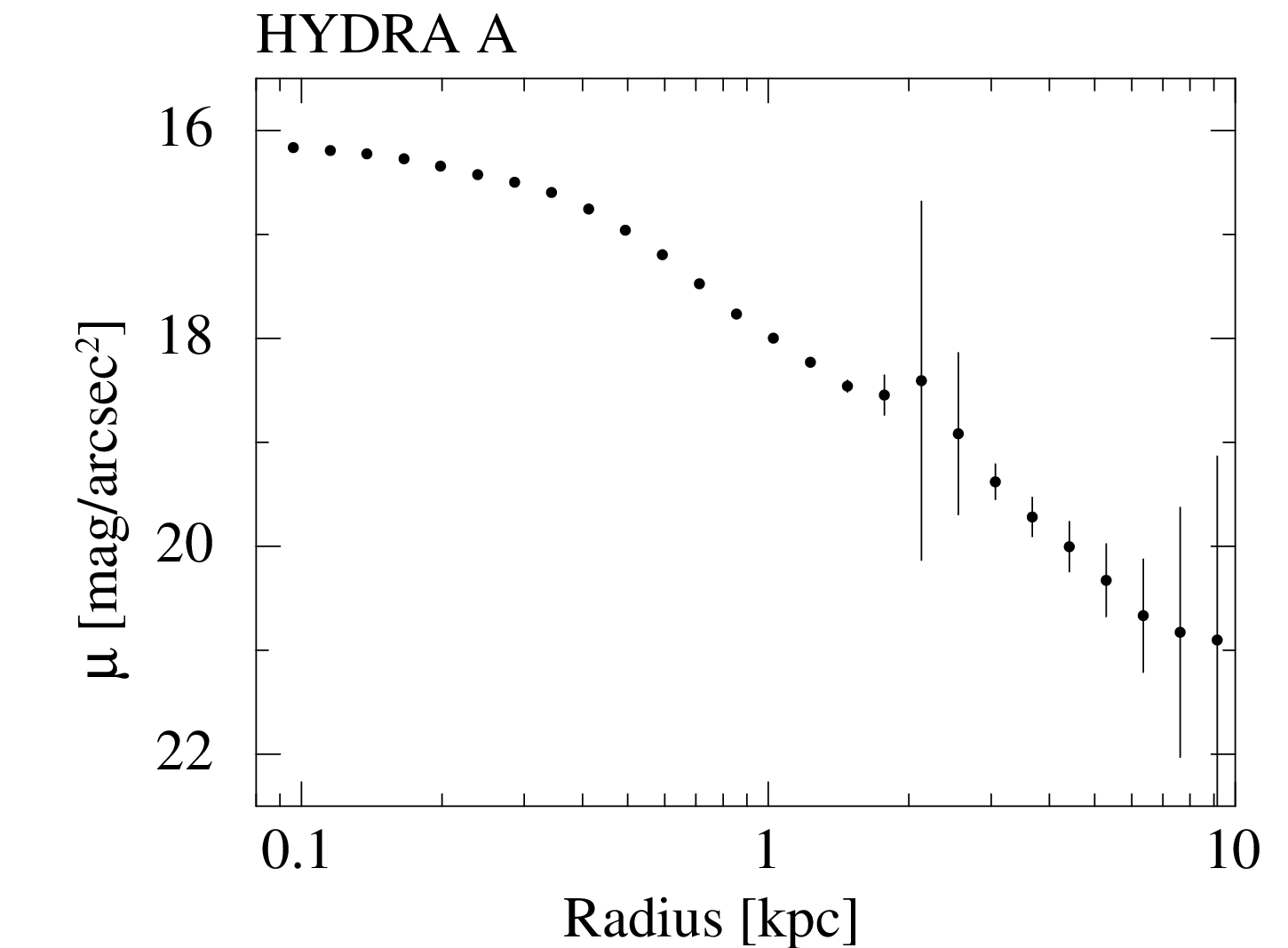}
   \includegraphics[width=6cm,angle=-90]{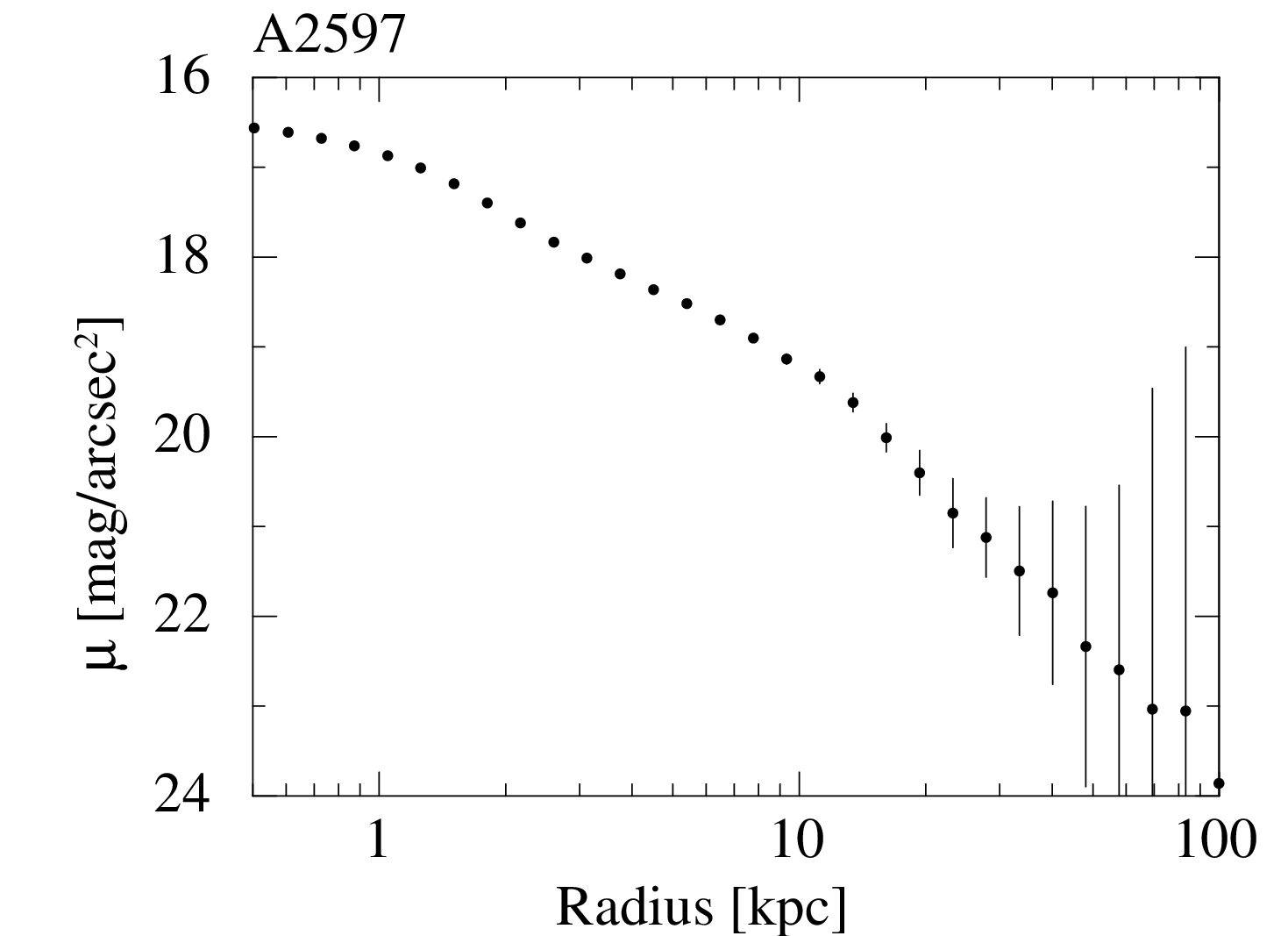}
   \includegraphics[width=6cm,angle=-90]{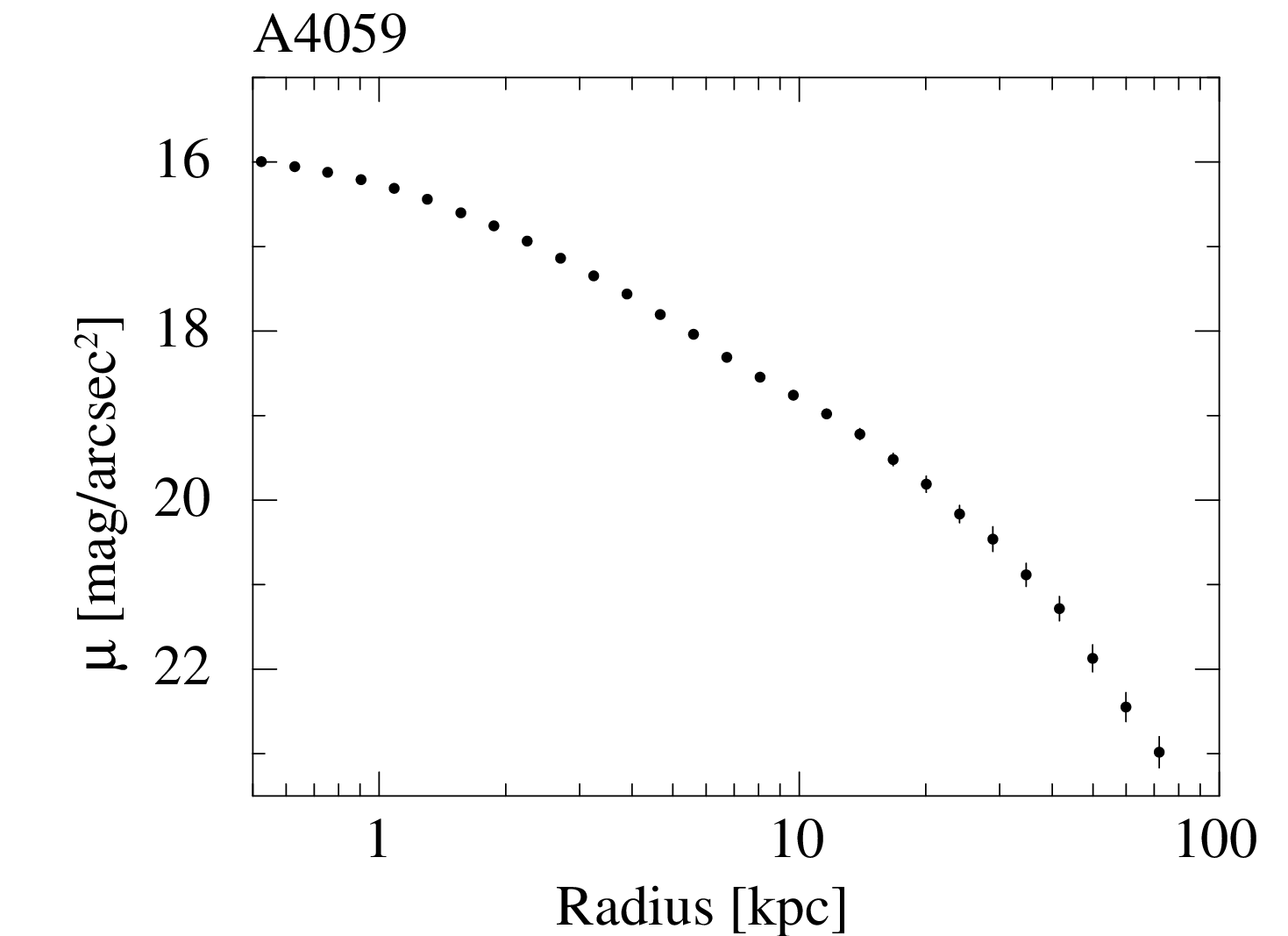}
   \includegraphics[width=6cm,angle=-90]{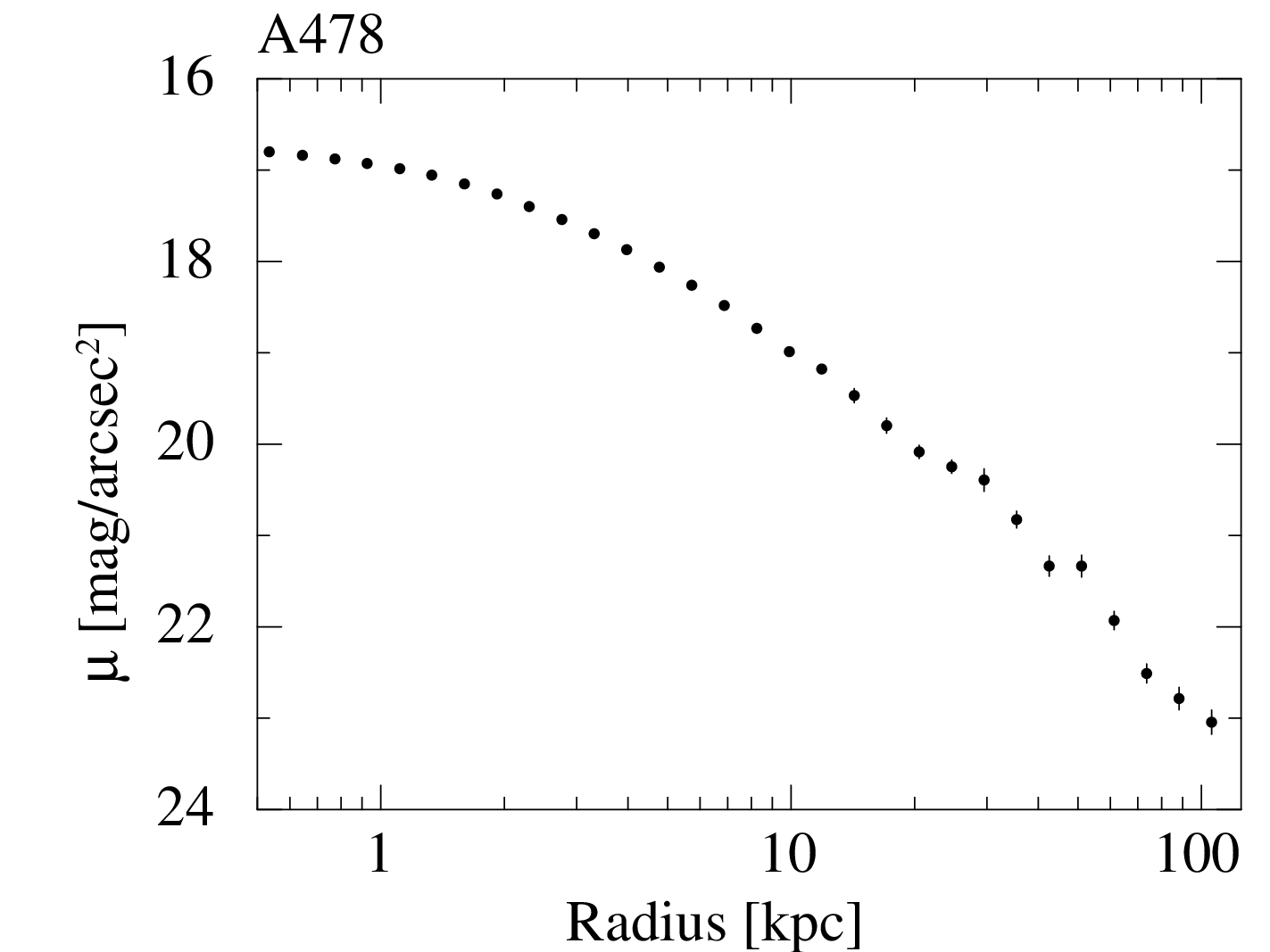}
   \includegraphics[width=6cm,angle=-90]{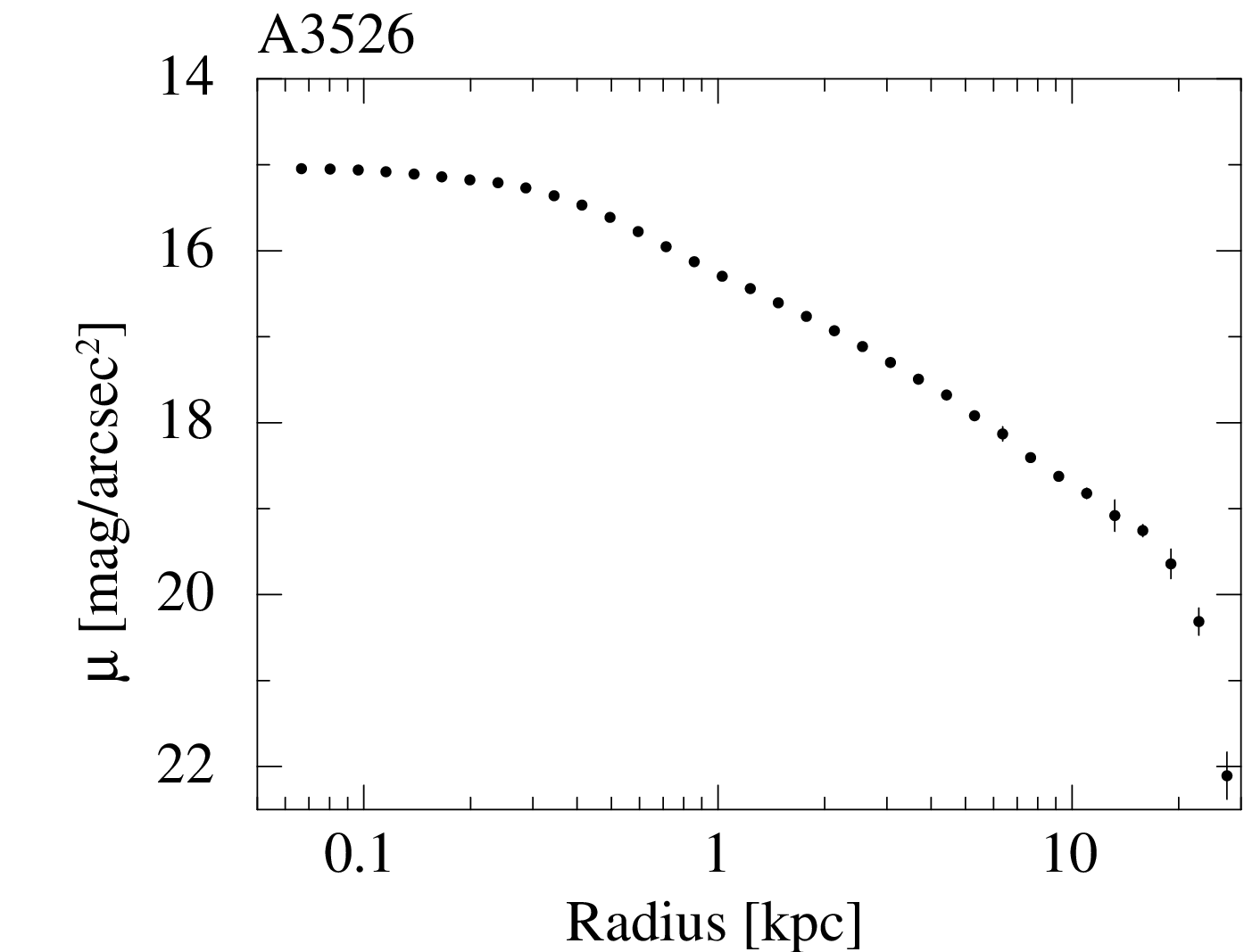}
   \includegraphics[width=6cm,angle=-90]{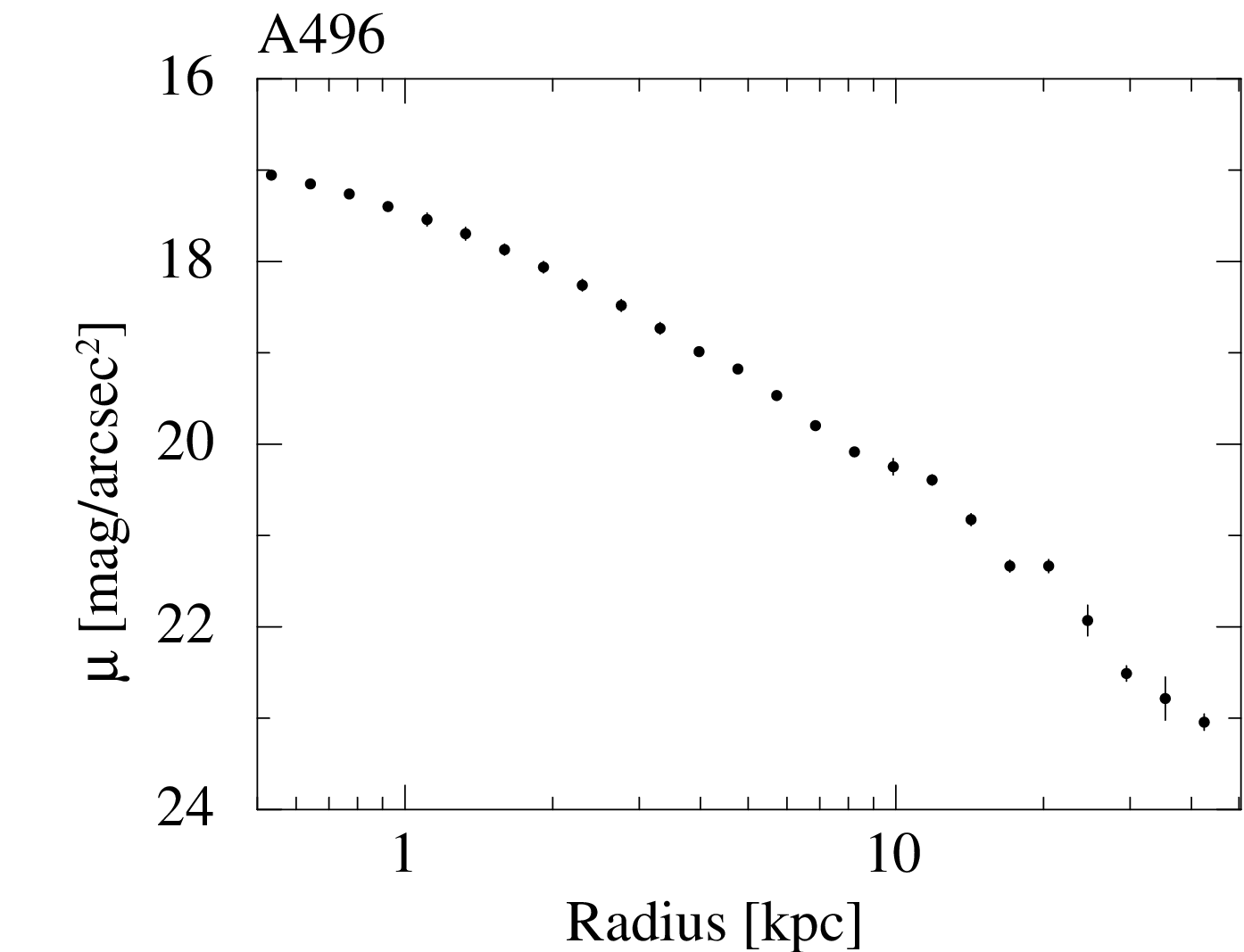}
    \caption{ Ks band surface brightness profiles of the BCGs in the
        local clusters listed in Table \ref{tab_loc}.
         }
         \label{fig_loc}
  \end{figure*}
%______________________________________________________________

\subsubsection{X-ray data}

We analysed XMM-Newton archival data for the clusters in
Table \ref{tab_loc}, following the data analysis described in details
in our previous work \citet*{degrandi09}.
We reprocessed the observation data files (ODF) using the Science
Analysis System (SAS) version 10.0.0.
The effective exposure times of the observations obtained after soft
proton cleaning are reported in Table \ref{tab_loc}.  Using the soft-proton, cleaned event files we extracted spectra from annular regions
centred on the cluster emission peaks.  We subtracted the background
using blank-sky fields for EPIC MOS and pn produced by
\citet{leccardi08_t}.
All the spectral fits were performed with the XSPEC package
using a one-temperature thermal
model ({\it apec}) and a two-temperature thermal model ({\it
  apec+apec}).
The Galactic hydrogen column is accounted for with a
{\it wabs} absorption model component assuming a Galactic HI value
by \citet{kaberla05}.
Redshifts, temperatures, iron
abundances, and normalizations of the two models were left free to vary.
Whenever the two-temperature fit provided significant improvements
over the single-temperature model (using an F-test), we used the Fe abundance
derived from the two-temperature model.

%----------------------------------------------------------------------
 \begin{table}
 \begin{center}
 \caption{Ks band observations parameters of the BCGs in the local
   clusters.  The columns are: (1) cluster name, (2) zero point in
   VEGA magnitudes (to convert from VEGA to AB magnitudes
   the correction is $+1.86$), (3) rms of the zero point for the night of
   measurement, (4) number of combined frames (each frame has 6 s
   exposure), and (5) seeing of image in arc seconds.}
 \label{tab_optloc}
 \begin{tabular}{l l l c c}                                             % centered columns (4 columns)
& & & &  \\
\hline\hline                                                                  % inserts double horizontal lines
Name          &    zp   & rms$_{\rm zp}$  & N  & Seeing(") \\
\hline
A478           &    22.301     & 0.132        & 50  & 0.81 \\
A496           &    22.316     & 0.016        & 30  & 0.63 \\
%A2029      &    22.884     & 0.0506      & 23  & 1.44 \\
A2597         &    22.373     & 0.028        & 30  & 0.69 \\
A3526         &    22.417     & 0.154        & 16  & 0.98 \\
%A4038      &    22.380     & 0.018        & 20  & 0.72 \\
A4059         &    22.252     & 0.141        & 23  & 0.81 \\
Hydra-A      &    22.269     & 0.057        & 38  & 0.75 \\
%2A0335+096 &     22.316     & 0.016           & 30              & 0.60 \\
\hline
\end{tabular}

 \end{center}
 \end{table}
%----------------------------------------------------------------------

\subsubsection{Near-IR data}

When deriving the NIR surface brightness profiles of the BCGs in the local
cool-core clusters, we used Ks band data that were acquired
using SofI \citep{moorwood98}
mounted  on NTT telescope at the ESO/La Silla observatory.
Since the targeted galaxies have an angular size ranging from 2 to 4
arcmin, the observations were done with the large field mode,
corresponding to a field of view of $5\times 5$ arcmin, with a pixel
scale of 0.288 arsec/pix.

We performed a standard data reduction of the raw frames and calibration
standard star fields using IRAF procedures.  The main steps of the
processing are dark subtraction, flat-field correction, and sky
subtraction.  Given the large angular size of the galaxies, sky
subtraction is achieved through an on-source/off-source imaging
strategy, where one frame targeted the science object and the next
observed a blank field.  Photometric calibration standards
\citep{persson98} were acquired several times during the observation
run.  The scatter in the zero points ranges from 0.02 to 0.14 mag (see
Table \ref{tab_optloc}).

The Ks band background subtracted surface brightness profiles
for the BCGs in the local cluster sample are shown in Fig. \ref{fig_loc}.
These Ks light profiles are integrated up to the background level
so they encompass faint light contribution, such as haloes
or intra-cluster light, up to measured radii.
For A3526, the CCD frame border is at $1.4^\prime$
($\sim$ 17 kpc) from the centre of the BCG, and therefore the
decreasing of the profile at larger radii is a border effect.
For this cluster we
limited our analysis to radii up to 17 kpc. In all other cases, we
verified that the CCD borders are well outside the range considered
in these plots.

We performed a 1D fit of the BCG light profiles shown in Fig. \ref{fig_loc}
using a Sersic model and integrated the best-fit model
up to a radius of 18 kpc to obtain the aperture Ks magnitudes of
the galaxies
(except for A3526 where we used the magnitude estimated by
\citealt{hoffer12}).
The absolute Ks magnitudes
(AB reference system) are reported in Table \ref{tab_mfe},
and errors on the magnitudes are $\pm0.10$ mag.

\subsection{A note on the NIR band}

To probe the mass distribution of the BCG in detail and to measure
possible faint contribution from intra-cluster light, we have based our
analysis on deep Ks band data rather than other filter passbands.
Indeed, NIR luminosities and, in particular, K-band magnitudes,
represent a more direct link with the stellar mass content of galaxies
than the visual ones \citep{kauffmann98}.
The light in the visual bands is dominated by the brightest
main-sequence stars, whose luminosity decreases as stellar populations
age. In contrast, a sort of conspiracy in the post main sequence
evolution of stars maintains the K band luminosity of a stellar
population roughly constant when it ages \citep{charlot96}.
The K-band magnitudes vs. X-ray-derived physical quantities
are expected to correlate with a
smaller scatter with respect to the optical ones, and
this will allow us to put stronger constraints on the link between the
BCGs and the ICM properties.

The rest frame Ks band at $z\sim$1 falls on the 4.5 $\mu m$
channel of Spitzer/IRAC (Infrared Array Camera).
However, IRAC does not have enough spatial resolution to effectively
estimate surface brightness profile for the BCG of WARPJ1415 at
$z=1.03$.
We thus applied a K correction to the measured Ks-band data,
based on a fit to the galaxy SED,
to compare the BCG magnitude with the local sample.
We have not considered any size or luminosity evolution for
the distant BCG.

The Ks profile of the distant BCG, shown in Fig.\ref{fig_optprof_bcg},
samples a slightly different rest frame with respect to the local BCGs,
but still in the NIR band (i.e. in the J band corresponding to
$\sim 1.2\mu m$), where the contamination from stellar evolution
of the brightest main-sequence stars is still negligible.

%----------------------------------------------------------------------
 \begin{table*}
 \begin{center}
 \caption{ICM Fe mass computed within overdensity $\Delta=2500$, Fe
   mass excess, and magnitude within an aperture of 18 kpc for the distant
   cluster WARPJ1415 and for the local XMM-Newton cluster sample.
 }
 \label{tab_mfe}
 \begin{tabular}{l c c c c }                                             % centered columns (4 columns)
& & & &  \\
\hline\hline                                                                  % inserts double horizontal lines
Name          &    $r_{2500}$   & $M_{Fe,2500}$        & $M_{Fe}^{exc}$   & M$_{\rm K}$         \\
                   &    kpc            & $10^{9}$ $M_{\odot}$ &  $10^{9}$ $M_{\odot}$  & mag   \\
\hline
%WARPJ1415    &  317.0    &   $0.76\pm0.13$   & $1.67\pm 0.40$  &  23.80  \\ old S12 value
WARPJ1415 &  317.0    &   $7.57\pm1.26$   & $1.67\pm 0.40$  &  -24.73  \\
\hline
A478            &    641.4     &  $29.6\pm 0.07$   & $7.56\pm 0.64$ & -22.50  \\
A496            &    461.0     &  $10.0\pm 0.01$   & $3.99\pm 0.13$ & -23.96  \\
A2597          &    465.2     &  $9.10\pm 0.10$   & $1.46\pm 0.44$ & -23.47  \\
A3526          &    439.3     &  $4.60\pm 0.01$   & $0.82\pm 0.01$ & -24.07  \\
A4059          &    473.2     &  $10.4\pm 0.05$   & $4.86\pm 0.45$ & -24.09  \\
Hydra-A       &    466.3     &  $8.40\pm 0.02$   & $1.30\pm 0.14$ & -24.04  \\
%Name          &    $r_{2500}$   & $M_{Fe,2500}$        & $M_{Fe}^{exc}$           & $M_{Rc}$ \\
%                   &    kpc            & $10^{10}$ $M_{\odot}$ &  $10^{9}$ $M_{\odot}$   &   $<18$kpc         \\
%\hline
%%WARPJ1415  &  317.0    &   $0.76\pm0.13$   & $1.67\pm 0.40$  &  23.80  \\ old S12 value
%WARPJ1415  &  317.0    &   $0.76\pm0.13$   & $1.67\pm 0.40$  &  24.97 \\
%\hline
%A478           &    641.4     &  $2.96\pm 0.07$   & $7.56\pm 0.64$  &  24.81 \\
%A496           &    461.0     &  $1.00\pm 0.01$   & $3.99\pm 0.13$  &  23.69 \\
%A2597         &    465.2     &  $0.91\pm 0.10$   & $1.46\pm 0.44$  &  24.81 \\
%A3526         &    439.3     &  $0.46\pm 0.01$   & $0.82\pm 0.01$  &  23.88$^{a}$ \\
%A4059         &    473.2     &  $1.04\pm 0.05$   & $4.86\pm 0.45$  &  25.38 \\
%Hydra-A      &    466.3     &  $0.84\pm 0.02$   & $1.30\pm 0.14$  &  24.03 \\
\hline
%\footnotesize{ $^a$ value from DM01.}
\end{tabular}

 \end{center}
 \tablefoot{(a) M$_{\rm K}$ for the Centaurus cluster is from \citet{hoffer12}. }
\end{table*}
%----------------------------------------------------------------------

\section {Iron mass and iron mass excess}

\subsection{The iron mass}

The iron mass in the ICM (in solar units), enclosed within a
sphere of radius $r^\prime$ from the cluster centre,
is given by the following integral:
\begin{equation}
M_{\rm Fe} (<r^\prime) = 4\pi A_{\rm Fe} m_{\rm H} {Z_\odot
\over M_\odot}~ \int_0^{r^\prime} Z_{\rm Fe}(r) ~n_{\rm H}(r)~ r^2 dr ,
\label{eq:mfe}
\end{equation}
where $A_{\rm Fe}$ is the atomic weight of iron, $m_{\rm H}$  the
atomic unit mass, and $Z_{\rm Fe} = n_{\rm Fe}/n_{\rm H}$ is the
de-projected Fe abundance in solar units. ($n_{\rm Fe}$ and $n_{\rm H}$
are the iron and hydrogen densities by number.)
We estimate the iron mass in WARPJ1415 within $r_{2500}$
(i.e. $\lesssim 317^{+22}_{-18}$ kpc, see Table 1 in S12)
linearly interpolating the cumulative
iron mass profile computed with Eq. (\ref{eq:mfe}).

The errors on this mass are dominated by the systematics
due to the large errors in the
Fe abundance profile of WARPJ1415 for radii $\gtrsim 0.2
r_{200}\sim 180$ kpc (S12),  $Z_{\rm  Fe}^{\rm out}$.
To estimate these systematic errors
we compute the Fe mass with Eq. (\ref{eq:mfe}) twice, fixing
$Z_{\rm  Fe}^{\rm out}$ to two extreme values:
the best-fitting value of WARPJ1415 de-projected iron
abundance profile at $r > 0.2 r_{200}$ and the iron abundance
outside the core measured in nearby cool-core clusters
\citep[DM01;][]{leccardi10}.
The former value 
($0.21\pm 0.16$) represents a lower limit for
the Fe abundance outside the core in WARPJ1415, and
it takes the conservative possibility of a positive
evolution of $Z_{\rm  Fe}^{\rm out}$ from $z=1$ to $z=0$ into account.
The latter value \citep[$0.34 \pm 0.01$, from][converted into \citealt{asplund09}
solar units]{leccardi10} represents an upper limit
to the Fe abundance in the cluster outskirts
or, in other words, assumes that $Z_{\rm  Fe}^{\rm out}$
at redshift 1 is the same as the one at $z< 0.3$ \citep{balestra07,leccardi10}.
We finally compute the mean between these upper and
lower Fe masses to derive the Fe mass within $r_{2500}$:
M$_{\rm Fe,2500} = (7.57\pm 1.26)\times 10^{9}$ M$_\odot$,
with the systematic errors computed as half the difference
between the two mass limits.

We de-project the XMM-Newton profiles of the local clusters in
Table \ref{tab_loc}  following the method described in \citet{ettori02}
and \citet{ghizzardi04} and estimate the Fe masses of these
clusters within $r_{2500}$ with Eq. (\ref{eq:mfe}).
We use the relation $r_{2500} \sim 0.3 r_{200}$
\citep[e.g.][see their Appendix]{ettori10} with $r_{200}$ taken from
\citet*{reiprich02}.
The Fe masses within $r_{2500}$ are given in Table \ref{tab_mfe}.
From this table and the results in DM01 (for clusters in
common between the two samples we take the new XMM-Newton values),
we find that the Fe mass distribution for cool-core clusters in the local
Universe ranges between
$\sim 0.4-6 \times 10^{10}$ M$_{\odot}$ and that the Fe mass
obtained for WARPJ1415 is consistent with the Fe masses measured
in nearby clusters.

\subsection{The iron mass excess}

We use Eq. (\ref{eq:mfe}) to also measure the iron mass {\it excess},
$M_{\rm Fe}^{\rm exc}$, in WARPJ1415 and in the local clusters.
This mass is computed by substituting in Eq. (\ref{eq:mfe}) $Z_{\rm Fe}$ with
$Z_{\rm Fe}^{\rm exc}$, which is defined as the iron abundance associated
with the abundance peak in the cluster core after subtracting the
average Fe abundance measured at radii larger than 0.2$r_{200}$:
$Z_{\rm Fe}^{\rm exc}(r) = Z_{\rm Fe}(r) - Z_{\rm Fe}^{\rm out}$.

We compute the Fe mass excess and its errors for WARPJ1415 taking the two limits discussed previously for $Z_{\rm Fe}^{\rm
  out}$ 
into account(i.e., 0.21 and 0.34), obtaining $M_{\rm Fe}^{\rm exc} =
(1.67\pm0.40)\times 10^9$ M$_\odot$. 
\footnote{In agreement with S12, who find
a Fe mass excess of $1.8^{+0.7}_{-0.5}\times 10^9$ M$_\odot$,
the small difference is due both to our slightly different spectral analysis
and to the integration routine used to calculate the Fe mass excess.}
This measure is consistent with the Fe mass excess distribution for
nearby galaxy clusters: $\sim$ 1.0--10 $\times 10^9$
M$_\odot$ (see Table \ref{tab_mfe} and DM01).
For WARPJ1415 we find a
ratio M$_{\rm Fe}^{\rm exc}$/M$_{\rm Fe}$ of $0.22\pm 0.06$, which
again agrees within the uncertainties with the local cool-core clusters
mean value, $0.20\pm 0.05$.

By comparing the Ks absolute magnitude of the BCG in WARPJ1415
with those of the local sample shown in Table \ref{tab_mfe}, we find
that their values are comparable.
This, and the similarity between the Fe mass excess in this and in local
systems \citep[see also][]{degrandi04}, strengthens our hypothesis
that the iron present in
the excess has also been produced in this distant cluster by the
stars that form the BCG at z=1.

\subsection {Iron abundance excess formation time scale}

The high redshift of WARPJ1415 allows us to put some constraints on
the formation time scale of the Fe peak.
We use the simple formalization given by \citet[][]{boehringer04b},

\begin{equation}
t_{enrich} = M_{\rm Fe}^{\rm exc}[L_B^{BCG} ({\rm SNR} ~10 ^{-12} L_{B\odot}^{-1})\eta_{\rm Fe}]^{-1}
\label{eq:ten}
\end{equation}
where $L_B^{BCG}$
is the luminosity in the B-band of the BCG,
SNR is the SN type Ia rate at
the redshift of the cluster measured in SNU (1 SNU $= 10^{-12}$ SNe
L$_{B \odot}^{-1}$ yr$^{-1}$),
and $\eta_{\rm Fe}$ is the iron yield per SN.
In Eq. (\ref{eq:ten}) the Fe mass excess is assumed to be
entirely produced by Type Ia SNe neglecting additional minor
contribution such as stellar mass loss or newly formed stars in the
BCG.
The rate of SNIa has been recently measured to be
$0.50^{+0.19}_{-0.23}$ SNU in galaxy clusters at $0.5 < z <1.46$
\citep{barbary12}. This value suggests that there is little or no
evolution in the cluster SNe rate from $z\sim 1$ to the present time.
\citep[For a compilation of recent cluster rate measurements
see][]{dilday10,maoz12r}.
With a standard yield of $M_{\rm Fe} = 0.7 M_\odot$ per SNIa
\citep{iwamoto99}, we find $t_{enrich} = 16.2$ Gyr, which is
significantly greater than the age of the Universe at the redshift of
WARPJ1415 (i.e. $t_{H(z=1)}=5.9$ Gyr).  We conclude that current
estimates of the Fe production rates at high redshift fail to account
for the presence of the high Fe abundance peak at $z=1$ in WARPJ1415.

%-----------------------------------------------------------
   \begin{figure*}
   \centering
   \includegraphics[width=6cm,angle=-90]{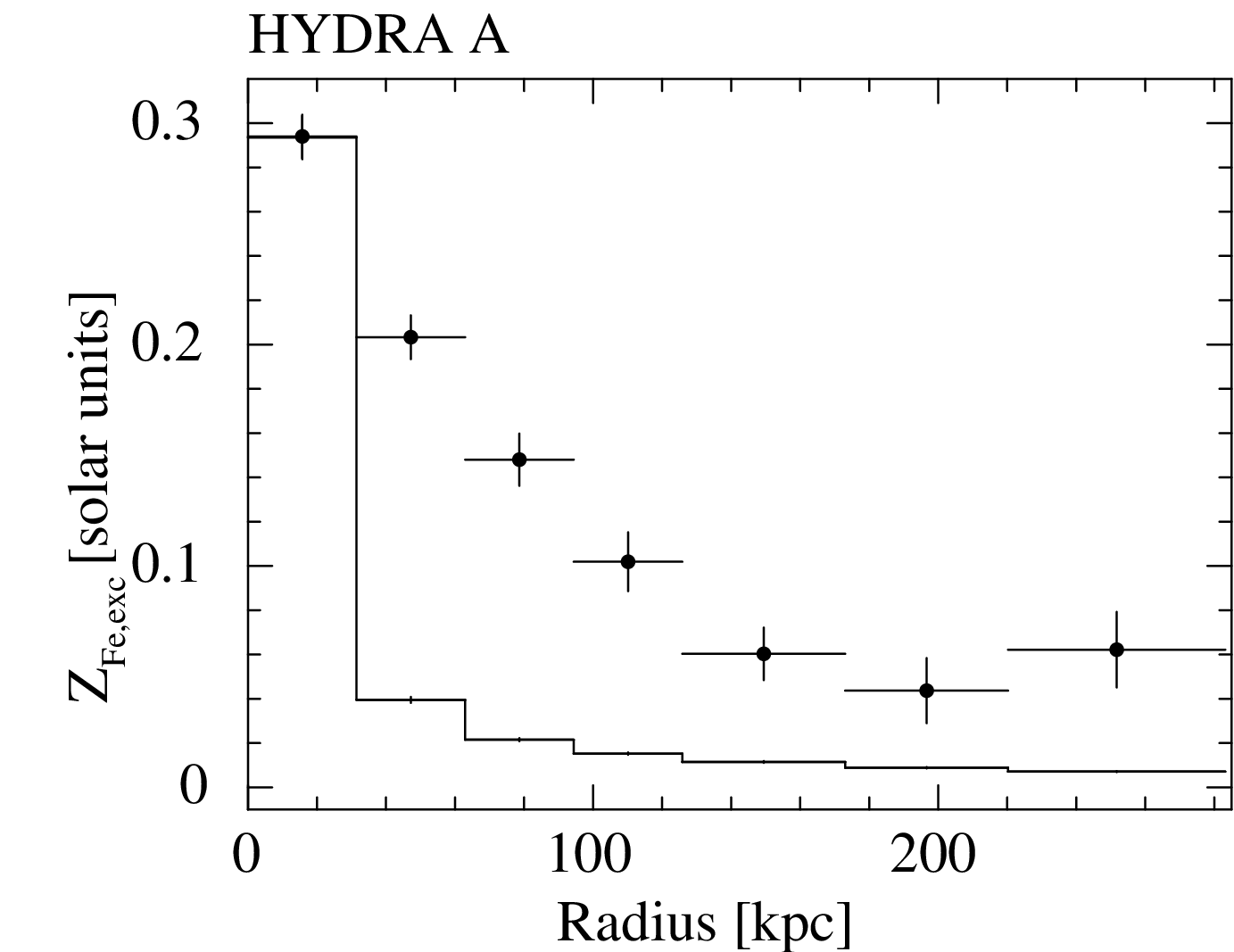}
   \includegraphics[width=6cm,angle=-90]{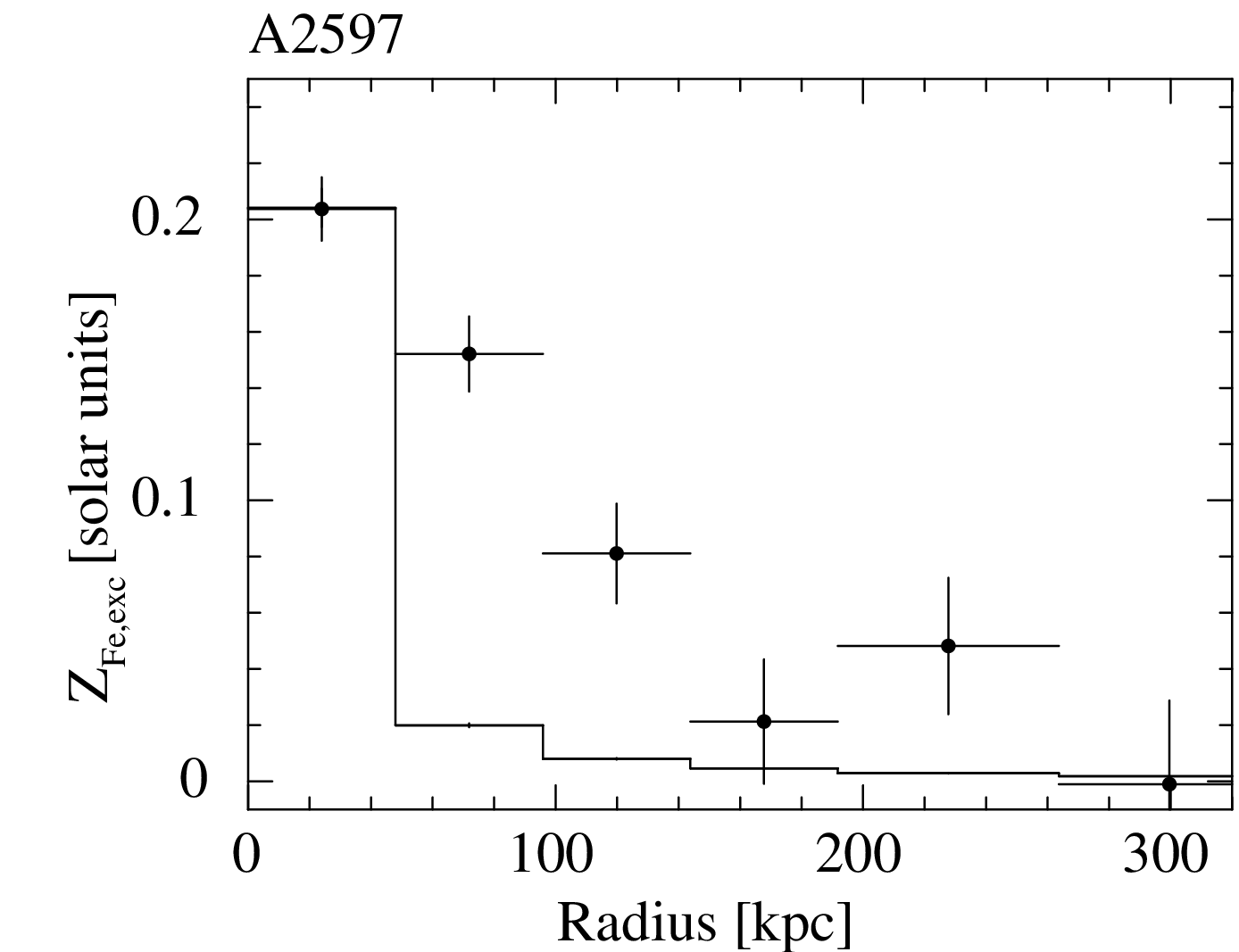}
   \includegraphics[width=6cm,angle=-90]{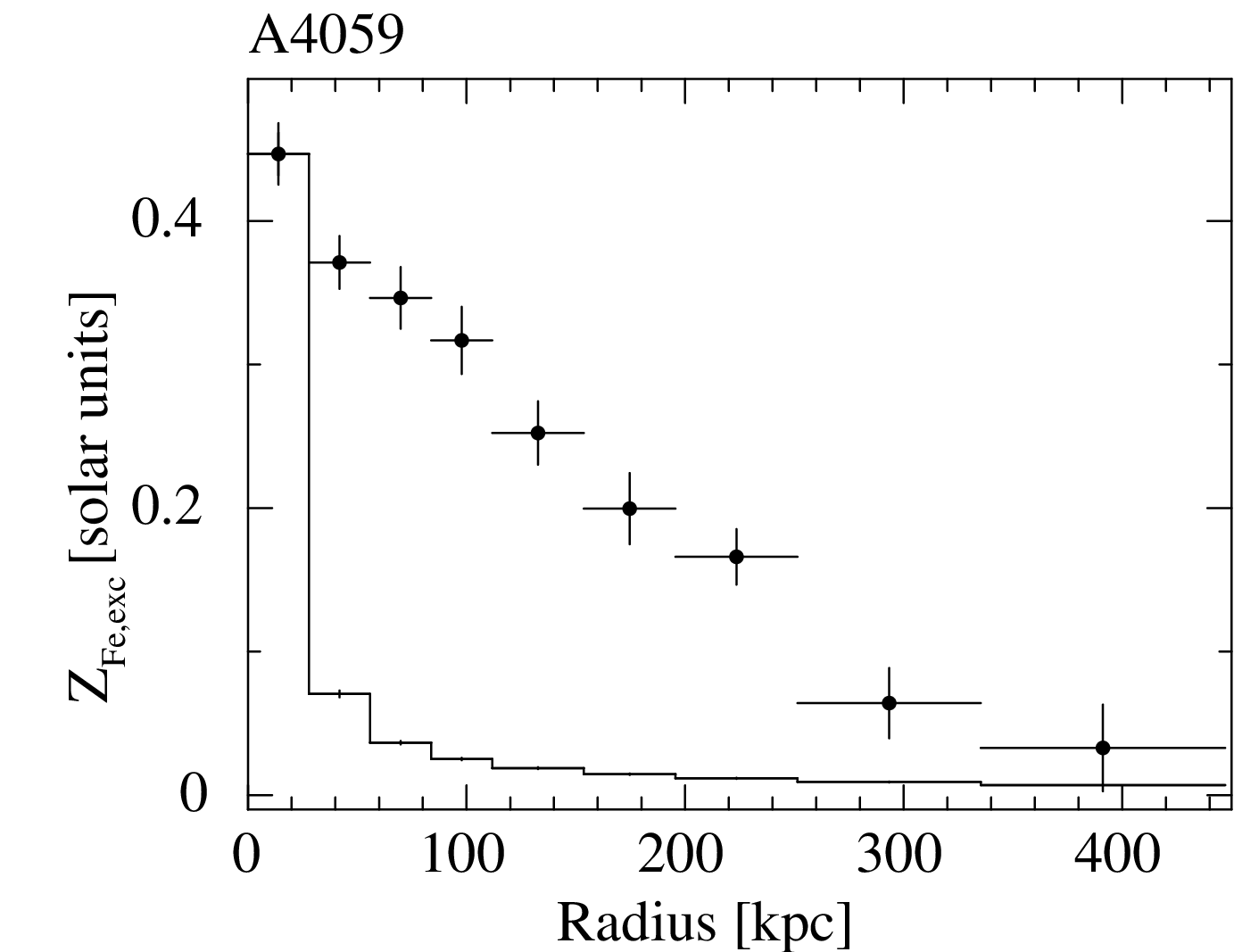}
   \includegraphics[width=6cm,angle=-90]{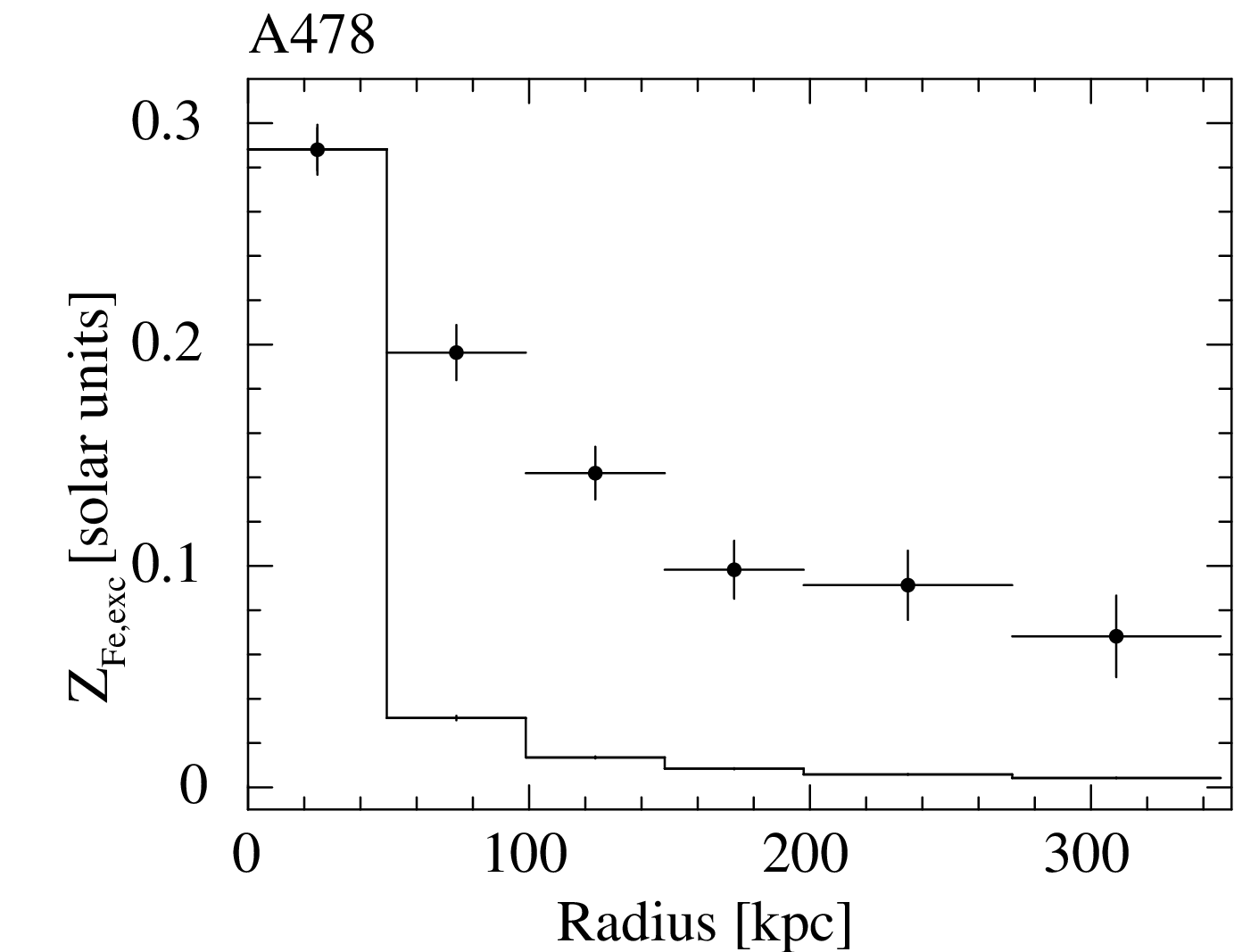}
   \includegraphics[width=6cm,angle=-90]{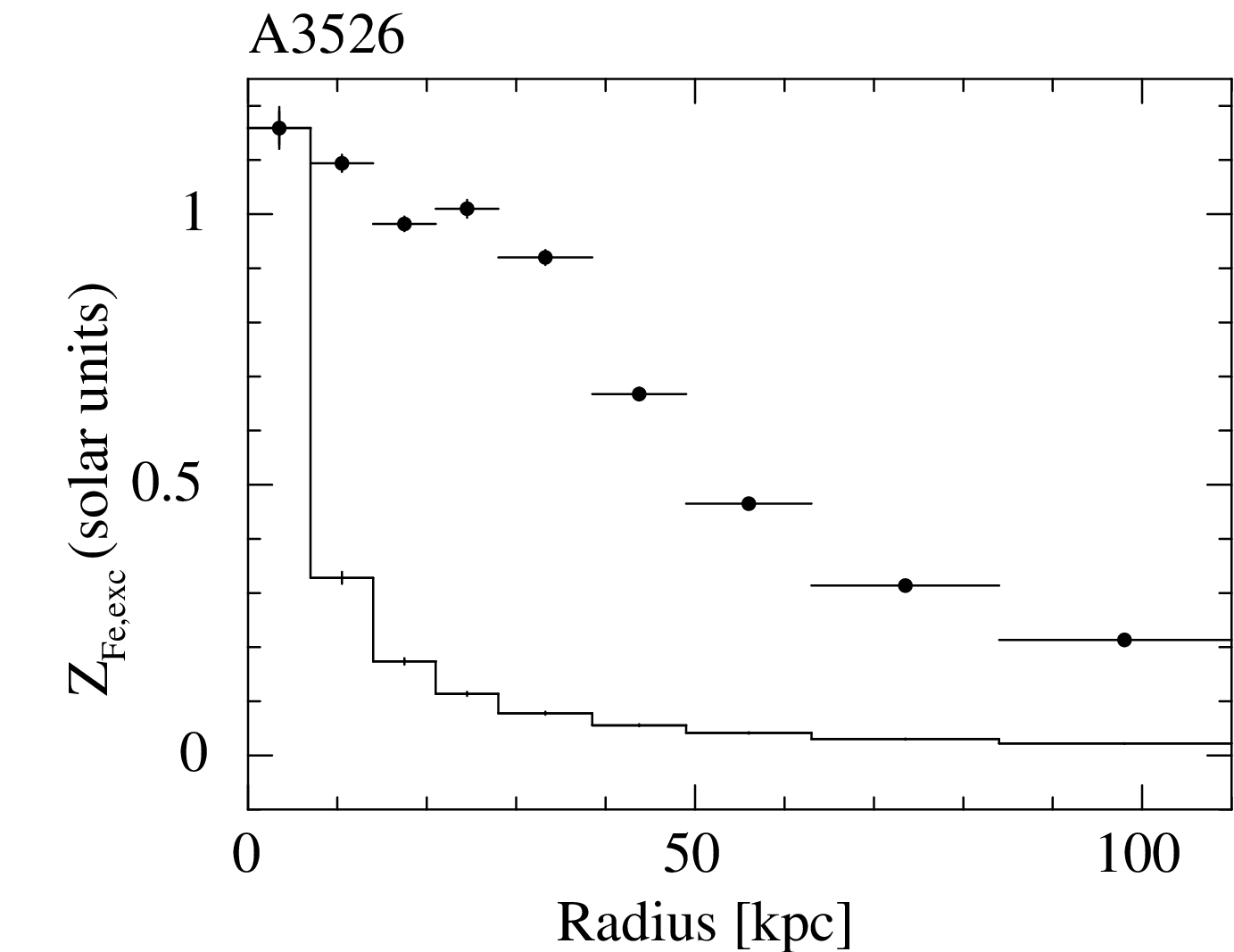}
   \includegraphics[width=6cm,angle=-90]{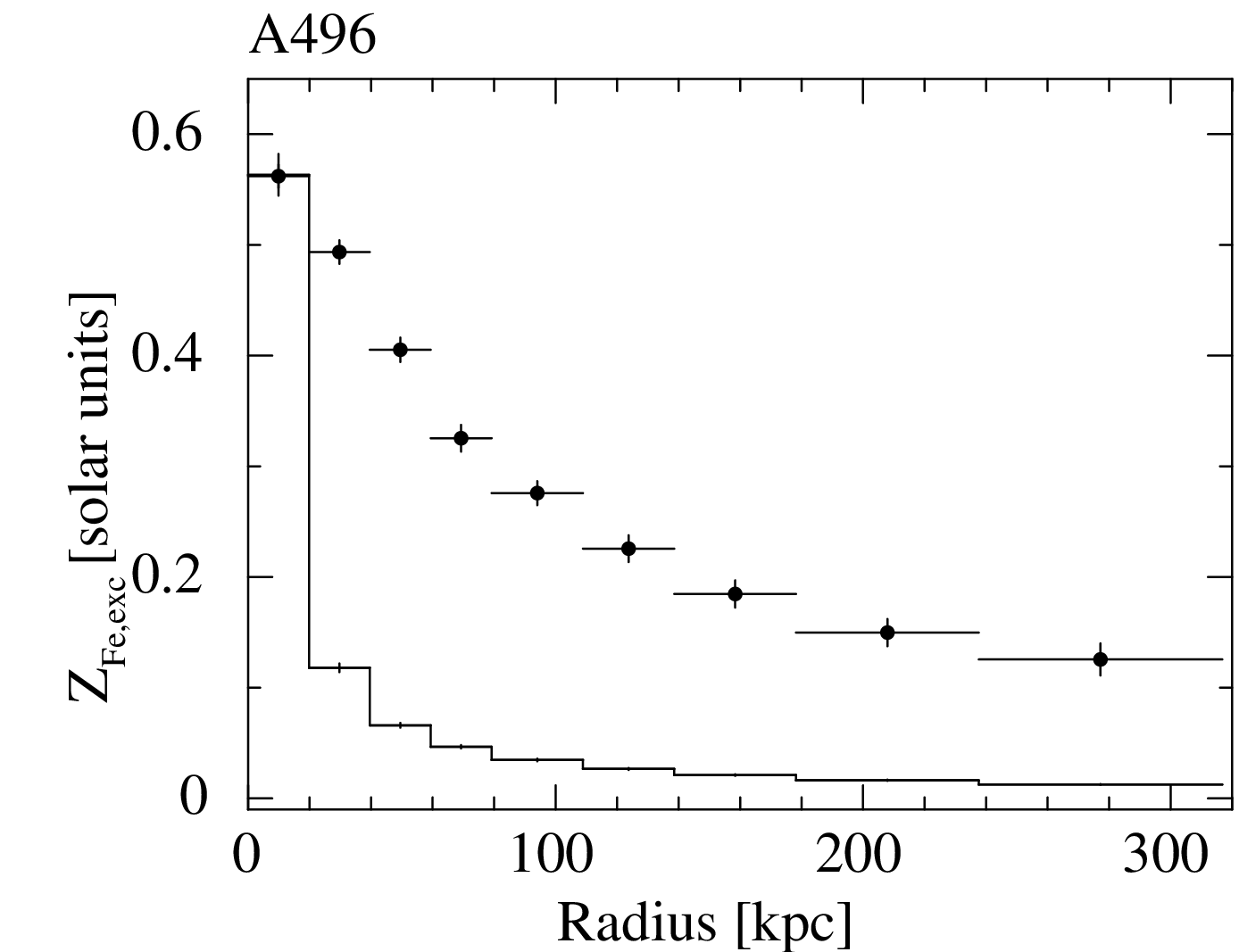}
     \caption{Comparison between the observed (black dots) and
       expected (red lines) Fe abundance excess profiles for the
       galaxy clusters in the XMM-Newton local sample.  }
         \label{fig_lz_loc}
   \end{figure*}
%______________________________________________________________

\subsection {Constraints on the SN rate in BCGs at $z>1$}

As shown in the previous section, a constant SNR is inconsistent with our results,
it is also inconsistent with other observational evidence and with theoretical work 
\citep[e.g.][]{maoz12r}. 
\citet{maoz10} argue that the SN delay time distribution (DTD)  can be described 
as a power law with index close to $-1$. To test that our data are consistent with 
such a model, we estimated the total number of SN that have exploded 
in the BCG by assuming a SNR of the form:
$ {\rm SNR} = {\rm SNR}_o \cdot (t/t_o)^{-1}$, 
where $t_o$ is the time that has passed between the formation of the BCG
and our observation of  WARPJ1415; SNR$_o$ is the SN rate at $t_o$, which we 
derive by multiplying the stellar mass of our BCG, $M_\ast$, with the SNR per unit 
mass (SNuM), i.e. ${\rm SNR}_o = {\rm SNuM} \cdot M_\ast$,
where we adopt the SNuM derived  by \citet{barbary12} for high-redshift
red-sequence early type galaxies (see their Table 7); and $M_\ast$ is taken 
from \citet{santos12}.
This leads to a value of SNR$_o = 0.474 \,{\rm yr}^{-1} $.
The total number of SNIa, ${\rm SN}_{tot}$, is obtained by integrating the SNR 
from a minimum delay time, $t_{min}$, to $t_o$, 
which leads to $ {\rm SN}_{tot} = {\rm SNR}_o \, t_o \, \ln(t_o/t_{min})$. 
Finally, the total mass of Fe produced is obtained by multiplying
the total expected SN, $ {\rm SN}_{tot} $,  by the yield per SN, $\eta_{\rm Fe}$. 
Other authors \citep[e.g.][]{greggio05} have proposed that at early times the SNR 
might fall off less rapidly, with a slope of $-1/2$. We have modelled the SNR in this 
scenario with a broken power-law model with index $-1/2$ from $t_{min}$ to a 
break time, $t_b$, and $-1$ from $t_b$ to $t_o$.
In this case the total number of SN, ${\rm SN}_{tot}$, turns out to be
${\rm SN}_{tot} = {\rm SNR}_o \, t_o \, \{ 2[1 - (t_{min}/t_b)^{1/2}] + \ln(t_o/t_b) \}$.

Before comparing the Fe masses that can be estimated from different choices 
of the model parameters, we note that these predicted masses should be compared 
with the total Fe mass, which is given by the sum of the Fe excess mass observed 
in X-rays and the Fe mass locked up in stars in the BCG.
The latter can be estimated from the total stellar mass assuming a typical metal 
abundance of $Z_{\rm Fe}  = 1.2 $, \citep[e.g.][and refs. therein]{maoz10}.
By performing this calculation, we  find an Fe mass locked up in stars in the BCG of
$2.1\times 10^9 M_\odot$ that, as expected, is close to the Fe excess
and a  total estimated Fe mass of  $3.9\times 10^9 M_\odot$, which can now be 
compared with expectations from the different DTDs.

We then start by considering the case of a DTD shaped like a broken power law  
with $t_{min} = 0.04$ Gyr and $t_b = 0.4$ Gyr and consider the two cases of BCG 
formation at $z=2$ and $z=3$ corresponding to $t_o$ of  2.2 and 3.6 Gyr, respectively. 
We find a total expected Fe mass of $1.8\times 10^9 M_\odot$ and 
$3.0 \times 10^9 M_\odot$, respectively, but both these values fall short
of the measured Fe mass in the case of a late BCG formation, $t_o = 2.2$ Gyr, 
by more than a factor of 2 and in the case of an early formation, $t_o = 3.6$, by about 30\%.
In the case of a power-law-shaped DTD with $t_{min} = 0.04$ Gyr, we find a total 
expected Fe mass of $2.3\times 10^9 M_\odot$ for $t_o = 2.2$ Gyr and 
$3.8 \times 10^9 M_\odot$ for $t_o = 3.6$ Gyr, respectively.
For the late BCG formation scenario, we are still below, by about a factor of 2, 
 the observed Fe mass, while for the early BCG formation scenario we are in 
good agreement.

Summarizing, our calculations suggest that a late BCG formation is unlikely, 
while an early BCG formation with a relatively flat SNR at early times, the 
broken power-law model, is below but possibly not inconsistent with the 
measured value. The best agreement is found with a combination of early 
BCG formation and steep SNR extending down to very short delay times, 
namely 0.04 Gyr. 

At the end of this section we can draw some conclusions.
The galaxy cluster WARPJ1415 at $z=1.03$, from its X-ray and NIR
properties, is virtually indistinguishable from a strong cool-core
cluster at $z\lesssim 0.1.$ Indeed, its Fe mass and  Fe mass excess,
its luminosity, and the stellar mass of its central BCG all
agree with what it is observed in local cool-core clusters.

The presence of a well-formed Fe peak at the centre of this distant
cool-core cluster, all similar to the ones found in nearby clusters,
places strong constraints on the formation time of the peak itself.
As shown previously, the enrichment time of the Fe
abundance peak is $\sim 4$ Gyr, and to produce all the Fe
 in the central excess and in the stars of  the BCG,
we require SNIa rates that are much higher than those observed in
both low-redshift %\citep[e.g.][and references therein]{dilday10}
or high-redshift galaxy clusters. %\citep[e.g.][]{barbary12} galaxy clusters.

The most straightforward implication of our results is that the bulk of
the enrichment occurred at very early times, within a few 100 Myr of the
formation of the BCG, possibly during the BCG assembly process.
Indeed, in the case of the power-law DTD, which best reproduces our data,
half of the SN explode within 0.4 Gyr of the formation of the BCG. This is
consistent with  observational evidence showing that the star formation rate
(SFR) during this time could have been
extremely high.
A good example is the so-called Spiderweb galaxy at
$z=2$ with an observed SFR of $1390 \pm 150$
M$_\odot$ yr$^{-1}$ \citep{seymour12}, which converts into an equally
huge SNR \citep[e.g.][]{calura07,maoz12}.
Other observations show that $\sim 50-60$ per cent of the
iron released by SNIa explosions occurs within $\sim 1$ Gyr of a
burst of star formation \citep[e.g.][and references therein for the
most recent measure of the delay-time distribution of SNe Ia as a
function of time]{maoz10,maoz12}.
Both observation \citep[e.g.][]{daddi04,rosati09,collins09,stott10} and theory
\citep[e.g.][]{dubinsky98,conroy07a,delucia07}, point out that BCGs form at
early times ($z>2$) relatively quickly via hierarchical merging (or cannibalism) of small
groups of galaxies falling along the filamentary large-scale structure
prior to cluster virialization.
The growth of the BCGs continue to be fed by mass flows up to
$z=1.5-2$, when the stellar population of the BCG start to evolve
mostly passively \citep[e.g.][]{brodwin13}.
Nearly all BCGs in the centres of local cool-core clusters still show little
ongoing star formation, with SFRs that correlate with the ICM
cooling rate \citep[e.g.][]{rafferty08,odea10,donahue10,mcdonald11};
however, these rates contribute to a negligible Fe production.
The SFR in WARPJ1415 of $\sim 2-8$ M$_\odot$/yr$^{-1}$ (S12),
as inferred from the [OII] emission line in the BCG spectrum,
is comparable to typical SFRs measured in BCG in cool-core clusters
at low redshift  \citep[i.e.,$1-10$ M$_\odot$/yr$^{-1}$][]{odea10}, indicating
that the major epoch of metal production in this cluster consistently
happened at redshifts higher than 1.

All these observations suggest that the bulk
of the Fe abundance excess in WARPJ1415 was likely produced by
the prompt release of the SNIa shortly after its BCG assembly at $z\gg1$
and relatively short delay times.

\section{BCG stellar light and Fe abundance profiles}

In the previous section we found that the Fe mass excess in
WARPJ1415 was most probably
produced by the stars forming the BCG at redshifts $\gtrsim 1.5-2$.
Now we want to compare the stellar light distribution of the BCG with
the Fe abundance excess profile.

To this aim we compute the Fe abundance profile {\it expected}
when the Fe distribution traces the light distribution of the BCG.
Following DM01, we find that the projected abundance $Z_{\rm Fe, proj}(b_{min},
b_{max})$ measured within a bin with bounding radii $b_{min}$,
$b_{max}$ is related to the de-projected Fe abundance distribution,
$Z_{\rm Fe}(r)$, by the following equation:

\begin{equation}
Z_{\rm Fe, proj}(b_{min},b_{max}) = { \int_{b_{min}}^{b_{max}} b~db
\int_{b^2}^{\infty} {n_H^2(r) Z_{\rm Fe}(r) \over \sqrt{r^2 -b^2} } dr^2
\over \int_{b_{min}}^{b_{max}} b~db
\int_{b^2}^{\infty} {n_H^2(r)\over \sqrt{r^2 -b^2} } dr^2 },
\label{eq:feexp}
\end {equation}
where $n_H(r)$ is the hydrogen density (by number), and
$Z_{\rm Fe}(r)=n_{\rm Fe}(r)/n_H(r)$ is the abundance excess profile
with $n_{\rm Fe}(r)$ equal to the Fe density (by number).
We assume that $n_{\rm Fe}(r)$ follows the stellar light distribution of
the BCG, $l(r)$, i.e.  $n_{\rm Fe}(r)\propto l(r)$, and then compute
the projected Fe abundance profile {\it expected}, $Z_{\rm Fe, proj}$, under this
hypothesis ($Z_{\rm Fe}(r) \propto l(r)/n_H(r)$) with Eq. (\ref{eq:feexp}).
We, finally, compare this {\it expected} Fe excess profile with the observed
one.

DM01 performed a similar comparison on a sample of local clusters
observed with BeppoSAX, showing that the observed Fe abundance
profile in the Perseus cluster was much broader than the
predicted one.  This result was later confirmed by several authors
working on nearby clusters observed with the XMM-Newton
satellite \citep[e.g.][]{churazov03,rebusco05,david08}.
Under our assumption that the iron in the excess
was formed by the stars in the BCG, this difference indicated that the Fe produced in
the BCG spread out several tens of kpc from the place of origin.

%-----------------------------------------------------------
   \begin{figure}
   \centering
   \includegraphics[width=6cm,angle=-90]{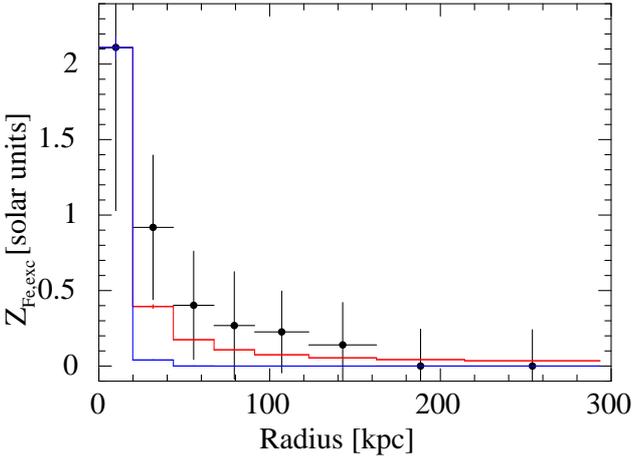}
      \caption{Comparison between the observed (black dots) and
        expected (lines) Fe abundance excess profiles for WARPJ1415 at
        $z=1.03$. The red and blue lines are the expected profiles
        computed taking the errors in the observed BCG's
        Ks profile shown in Fig. \ref{fig_optprof_bcg} into account.  }
         \label{fig_lz_dist}
   \end{figure}
%______________________________________________________________

We redid this exercise for the local cool-core clusters
listed in Table \ref{tab_loc} and for WARPJ1415 at $z=1$.  The results
are plotted in Fig. \ref{fig_lz_loc} for the local clusters and
in Fig. \ref{fig_lz_dist} for WARPJ1415: in each panel the data points
are the observed Fe abundance excess, whereas the continuous lines are
the expected Fe abundance excess profiles.
The profiles are arbitrarily normalized to the
innermost bin because in this context, we are only interested in the
comparison between the shapes of the profiles.
From Fig. \ref{fig_lz_loc} it is clear that for all the local clusters,
the observed profiles are substantially broader than the expected ones.

In Fig. \ref{fig_lz_dist}, we plot two expected
profiles for WARPJ1415 to include the uncertainties in the Ks band BCG's
profile (see Fig. \ref{fig_optprof_bcg}).
These uncertainties have been included by computing the maximum and
minimum expected Fe abundance excess profiles with Eq. (\ref{eq:feexp})
using the Ks band profile twice, the first time adding
its $1\sigma$ statistical errors to the Ks band profile and the second time,
subtracting it (Fig. \ref{fig_lz_dist}).
Since the errors in the cumulative Ks band profiles
are not independent, these systematic errors
may be overestimated.
We have also experimented with different spatial binnings of the
X-ray Fe abundance profile and find consistent results.
We conclude that, despite the larger scatter produced by the X-ray and
NIR measurements in WARPJ1415, its observed Fe abundance excess
profile is closer, or at most it differs marginally from the predicted
one.  This is at variance with what it is observed in low-redshift
cool-core clusters, where the Fe abundance excess profiles are
always significantly broader than the expected ones
(Fig. \ref{fig_lz_loc}).

Under the reasonable assumption that WARPJ1415 will evolve into a
present-day cool-core cluster, the smaller difference between the
observed and expected profiles (Fig. \ref{fig_lz_dist}) may indicate
that the processes responsible for the broadening of the Fe excess
profile in local systems have not yet had time to produce significant
modifications in this distant cluster.

The time available for widening the Fe abundance peak excess is
$\gtrsim 7.9$ Gyr, i.e., the look-back time at $z=1.03$.
On the basis of this observation alone, it is not possible to know
whether there is a dominant mechanism that broadens the Fe
profile or whether this widening is the result of the combined action of
more mechanisms. %,
Neither it is possible to hypothesize
whether the Fe broadening is the result of many smaller events during
a long period of time or if it is the result of a single event happening in
a relatively short time period. %,
What is apparent is that the time available
to broaden the Fe peak is very long.

The mechanisms that could mix the ICM and produce the observed
Fe profiles are several \citep[see][and references
therein]{schindler08}. %Schindler \& Diaferio 2008, ).
One possible mechanism is turbulent mixing of
the ICM due to AGN activity \citep[e.g.][]{rebusco05,simionescu09,simionescu10,gitti11,osullivan11,mcnamara13}.
\citet{kirkpatrick11} found that clusters where the BCGs have
experienced recent AGN activity
(observed in the form of bright radio emission, cavities, and shock
fronts embedded in the ICM)
have heavy elements distributed anisotropically and aligned with the
large-scale radio and cavity axes, indicating that they are being
transported from the halo of the BCG into the ICM along the
large-scale outflows driven by the radio jets.
The amount of transported gas is substantial, and it is consistent
with the results of simulations showing that AGN outflows are able to
advect ambient, iron-rich material from the core to much larger radii of hundreds of kpc \citep{gaspari11a,gaspari11b,planelles13}.

Another mechanism at work could be the ICM-sloshing induced by minor
mergers in cool-core clusters \citep{ascasibar06,roediger11},
a phenomenon that is now more frequently discovered in nearby cool-core
clusters \citep{simionescu10,ghizzardi10}.
Indeed, in \citet{ghizzardi13} the observations of the central regions of A496
show that the gas-sloshing is able to shift the metals up to several
tens of kpc out from the BCG \citep[see also][]{osullivan14}.

Although still debated, the stellar mass of the
BCGs appears to increase by a factor $\sim 2-3$ from redshift 1 to 0
(\citealt[][see however Collins et al. 2009, Stott et al. 2010]{lin13,laporte13,lidman12}).
This mass accretion occurs through mergers due to fast and close encounters
of small groups of galaxies onto the BCGs \citep[e.g.][]{lidman13,tonini12}.
The fate of the  merging galaxy stars, which are collisionless, is to
accrete onto the central giant galaxy increasing the stellar mass and the
intra-cluster light surrounding such a galaxy
\citep[e.g.][]{conroy07b,murante07,henriques10}. Indeed, once the
streams of stars have been formed, they begin to decay since they are disrupted by the tidal field
of the cluster within timescales that are $\simeq 1.5$ times their dynamical time
in the cluster \citep[$t_{dyn}\lesssim1$ Gyr,][]{rudick09}.
Less  obvious is the fate of the enriched ISM and of the infalling galaxies, which
behave as a collisional fluid.
This gas is probably tidally or ram-pressure stripped \citep{kapferer07}
from its disrupted galaxy and mixed with the ICM at a greater distance from the
BCG, with respect to the stars, in the end possibly contributing to the widening
of the Fe peak between redshift 1 and 0.
The relative importance of this process should be investigated in depth both
observationally and with numerical simulations \citep[see][]{vazza10}.

\section{Summary}

Spatially resolved studies of the iron content in high-redshift
clusters are very important because they provide much
needed information on the enrichment process in these systems.
Unfortunately, thus far, only a handful of  high redshift cluster
have been observed at the necessary sensitivity level, amongst them,
the X-ray luminous cool-core cluster WARPJ1415 at $z=1.03$, which
is the subject of this paper.
Our main findings can be summarized as follows.

\begin{enumerate}
\item By comparing the central Fe mass excess from WARPJ1415
with those of local cool-core systems, we find that the Fe abundance
excess produced by the BCG was already in place at $z=1$.
\item The large amount of metals observed at $z=1$ implies that
the enrichment process must have occurred early on.
More specifically, since SN rates at $z=1$ would imply enrichment
timescales that are longer than the age of the Universe,
our measures point to an early and intense period of star formation
most likely associated with the formation of the BCG.
\item The Fe mass excess in WARPJ1415 agrees with expectations
from the combination of early BCG formation (z $\sim 3$) and steeply
increasing
SNR in the past extending down to very short, i.e. 0.04 Gyr, delay times.
\item While for local cool-core clusters the Fe distribution is
broader than the NIR light distribution of the BCG, in
WARPJ1415 the two distributions are consistent, indicating
that the process responsible for broadening the Fe distribution
in local systems has not yet started in this distant cluster.
\end{enumerate}

\begin{acknowledgements}

The authors thank S. Borgani, S. Ghizzardi, and F. Mannucci for useful comments
that helped to improve this work.
SD, PT, SM, MR, and PR acknowledge the support by
INAF PRIN 2012 (P.I. S. Molendi),
AF acknowledges the support by INAF PRIN 2010 (P.I. L. Guzzo).
JS has received funding from the European Programme FP7/2007-2013
under grant agreement 267251 "Astronomy Fellowships in Italy" (Astrofit).
This research has made use of the NASA/IPAC Extragalactic Database
(NED), which is operated by the Jet Propulsion Laboratory, California
Institute of Technology, of the NASA's High Energy Astrophysics
Science Archive Research Center (HEASARC), and, of the XMM-Newton
archive.

\end{acknowledgements}

\bibliographystyle{aa}              % style aa.bst
\bibliography{biblio_Jan2014}  % your references Yourfile.bib

\end{document}